\documentclass[acmlarge]{acmart}

\usepackage{subcaption}
\usepackage{float}
\usepackage[normalem]{ulem}

\usepackage{algorithm,algcompatible,amsmath}
\algnewcommand\INPUT{\item[\textbf{Input:}]}%
\algnewcommand\OUTPUT{\item[\textbf{Output:}]}%

\AtBeginDocument{%
  \providecommand\BibTeX{{%
    \normalfont B\kern-0.5em{\scshape i\kern-0.25em b}\kern-0.8em\TeX}}}

\setcopyright{none}
\renewcommand\footnotetextcopyrightpermission[1]{} 
\begin{document}

\title[TelecomTM: A Fine-Grained and Ubiquitous Traffic Monitoring System Using Pre-Existing Telecom Cables as Sensors]{TelecomTM: A Fine-Grained and Ubiquitous Traffic Monitoring System Using Pre-Existing Telecommunication Fiber-Optic Cables as Sensors}

\author{Jingxiao Liu}
\email{liujx@stanford.edu}
\affiliation{%
  \institution{Stanford University}
  \city{Stanford}
  \state{CA}
  \country{USA}
  \postcode{94305}
}
\author{Siyuan Yuan}
\email{syyuan@stanford.edu}
\affiliation{%
  \institution{Stanford University}
  \city{Stanford}
  \state{CA}
  \country{USA}
  \postcode{94305}
}
\author{Yiwen Dong}
\email{ywdong@stanford.edu}
\affiliation{%
  \institution{Stanford University}
  \city{Stanford}
  \state{CA}
  \country{USA}
  \postcode{94305}
}
\author{Biondo Biondi}
\email{biondo@sep.stanford.edu}
\affiliation{%
  \institution{Stanford University}
  \city{Stanford}
  \state{CA}
  \country{USA}
  \postcode{94305}
}

\author{Hae Young Noh}
\email{noh@stanford.edu}
\affiliation{%
  \institution{Stanford University}
  \city{Stanford}
  \state{CA}
  \country{USA}
  \postcode{94305}
}

\renewcommand{\shortauthors}{Liu, et al.}

\begin{abstract}

We introduce the \emph{TelecomTM} system that uses pre-existing telecommunication fiber-optic cables as virtual strain sensors to sense vehicle-induced ground vibrations for fine-grained and ubiquitous traffic monitoring and characterization. {Here we call it a virtual sensor because it is a software-based representation of a physical sensor.} Due to the extensively installed telecommunication fiber-optic cables at the roadside, our system using redundant dark fibers enables to monitor traffic at low cost with low maintenance. Many existing traffic monitoring approaches use cameras, piezoelectric sensors, and smartphones, but they are limited due to privacy concerns and/or deployment requirements. Previous studies attempted to use telecommunication cables for traffic monitoring, but they were only exploratory and limited to simple tasks at a coarse granularity, e.g., vehicle detection, due to their hardware constraints and real-world challenges. In particular, those challenges are 1) unknown and heterogeneous properties of virtual sensors and 2) large and complex noise conditions. To this end, our \emph{TelecomTM} system first characterizes the geographic location and analyzes the signal pattern of each virtual sensor through driving tests. We then develop a spatial-domain Bayesian filtering and smoothing algorithm to detect, track, and characterize each vehicle. Our approach uses the spatial dependency of multiple virtual sensors and Newton's laws of motion to combine the distributed sensor data to reduce uncertainties in vehicle detection and tracking. In our real-world evaluation on a two-way traffic road with 1120 virtual sensors, \emph{TelecomTM} achieved 90.18\% vehicle detection accuracy, 27$\times$ and 5$\times$ error reduction for vehicle position and speed tracking compared to a baseline method, and $\pm$3.92\% and $\pm$11.98\% percent error for vehicle wheelbase and weight estimation, respectively.

\end{abstract}



\maketitle

\section{Introduction}

A traffic monitoring system, which automatically and continuously detects, tracks, and characterizes vehicles in moving traffic, is important for urban management, maintenance, and planning. For instance, a traffic monitoring system can track and predict traffic patterns to help reduce traffic congestion~\cite{1263039,mandhare2018intelligent,s16020157} and manage safety and emergency situations~\cite{kherraki2022deep,ro2007lessons,goel2012intelligent}. Also, with a detailed understanding of individual vehicle characteristics (e.g., vehicle number, size, weight) and the use of roads and bridges, we can image the near-surface (tens of meters under the ground) seismic properties of urban areas~\cite{Yuan2021,yuan2020near}, monitor critical transportation infrastructure~\cite{lydon2016recent,wang2004overview,10197_4867,doi:10.1080/15732479.2017.1415941} and efficiently determine the needs of future transportation projects~\cite{HUANG2021136,guide2001traffic,won2020intelligent}.

There are several existing technologies for traffic monitoring, such as vision-based systems~\cite{5309837,7458203,REINARTZ2006149} and pavement sensing systems (e.g., inductive loops~\cite{jain2019review,6957957,doi:10.3141/1719-14}, piezoelectric sensors~\cite{zhang2015new,li2006application,jain2019review}, and fiber optic sensors~\cite{tekinay2022applications,yuksel2020implementation}). However, these systems bring several drawbacks: Vision-based systems are perceived as privacy-invasive and are sensitive to reduced visibility caused by weather conditions. Pavement sensing systems only capture traffic information at specific locations as they are point sensing. Due to the high cost in installations and maintenance, it is difficult to scale up and achieve fine-grained monitoring using existing vision-based and pavement sensing systems. Furthermore, crowd-sensing approaches that use mobile phone data from the drivers/passengers~\cite{7079458,doi:10.1080/01441640500361108,zhong2021metatp} have been developed to enable cost-effective and high-resolution traffic monitoring. However, they only capture the vehicle position and speed information and have also raised privacy concerns.

To this end, we introduce the \emph{TelecomTM} system that uses pre-existing roadside telecommunication (telecom) fiber-optic cables as virtual strain sensors to sense vehicle-induced ground vibrations for fine-grained and ubiquitous traffic monitoring and characterization. In particular, \emph{TelecomTM} achieves vehicle detection, tracking, speed and position estimation, weigh-in-motion, and wheelbase estimation. Our system is based on the distributed acoustic sensing (DAS) technology, specifically, the Phase Sensitive Optical Time Domain Reflectometry ($\phi$-OTDR)~\cite{rao2021recent,muanenda2018recent}. {A DAS channel is called a ``virtual sensor" because it is a software-based representation of a physical sensor. It uses the readings of the backscattered light in the fiber to calculate the strain responses of the telecom cable.} \emph{TelecomTM} is built on the idea that vehicle motion creates unique vibration patterns on the ground and near-surface structures. When vehicles move on the road, the ground deforms due to the vehicles' self-weight~\cite{jousset2018dynamic,yuan2020near}. When vehicles pass by a structure (e.g., a roadway structure), forces applied by their wheels induce the structure to vibrate~\cite{doi:10.1177/14759217221081159,liu2020damage,yang2004vehicle,liu2020diagnosis}. These vibrations carrying information about vehicle characteristics (e.g., size and weight) are transmitted to the roadside telecom fiber conduits coupled to the earth and road structures. Our \emph{TelecomTM} system senses these vehicle-induced telecom fiber vibrations to monitor traffic ubiquitously and infer vehicle activities with fine-grained spatial resolution. 

\emph{TelecomTM} is a scalable, efficient, and cost-effective system. There are millions of kilometers of telecom fiber cables deployed around the world that can be utilized for ubiquitous traffic monitoring. For instance, in 2017, the length of the optical fiber cable network in China alone was more than 37 million kilometers~\cite{Statista}. Most telecom infrastructure utilizes pipes and conduits several meters under the ground and along the roadways to distribute around the urban area, which can capture fine-grained traffic information. It only requires connecting an optoelectronic instrument called the interrogator unit to one end of the fiber. The interrogator unit used by \emph{TelecomTM} can record strain data from a telecom fiber cable up to 100 km long in a high spatial-temporal resolution (up to 250 Hz and 1-meter channel spacing)~\cite{Yuan2021,liu2022vibration}. In addition, by taking advantage of unlit dark fibers (i.e., fibers that are not used for data transmission), \emph{TelecomTM} can continuously record data for years without any interference to regular telecommunication signals or any on-site sensor installation and maintenance~\cite{Lindsey2020,ajo2019distributed}.

Researchers have explored the idea of using dark fibers for traffic monitoring before; however, due to hardware constraints and real-world challenges, existing works are limited to exploratory simple tasks at coarse-granularity, e.g., vehicle detection and traffic speed estimation~\cite{narisetty2021overcoming,opta_tm,Yuan2021,VandenEnde2022a,yuan2022spatial,wang2021ground}. For example, incoherent OTDR~\cite{healey1986instrumentation} only has intensity measurement at a lower spatial resolution (e.g., 10-meter channel spacing). Importantly, existing works lack a systematic approach to cope with real-world challenges, which prevented them from achieving fine-grained traffic monitoring and individual vehicle characterization. In particular, the key research challenges are:
\begin{itemize}
    \item {\bf Unknown and heterogeneous properties of virtual sensors}. \emph{TelecomTM} measures the dynamic strain of the fiber around each virtual sensor. Due to cable spooling in cabinets and manholes, each virtual sensor's actual geographic location (geo-location) is unknown. Therefore, the surrounding conditions (e.g., coupling between the cable and the conduit and between the conduit and the earth, near-surface soil properties, etc.) of each virtual sensor is unidentified, resulting in high uncertainties and heterogeneity in signal properties, including signal pattern and the ratio between the virtual sensor's response and vehicle-induced forces (i.e., transmissibility). Without a prior understanding of the geo-location and signal properties, we cannot accurately model vehicle-induced telecom fiber responses at each virtual sensor. Vehicle detection and tracking could also fail due to inaccurate vehicle position and speed estimations. 
    \item {\bf Large and complex noise conditions}. Fiber cables were originally deployed for data transmission as opposed to strain sensing. Telecom fiber responses are indirect measurements of the vehicle-induced ground vibrations, which have larger and more complex noises and uncertainties than direct measurements from dedicated and well-calibrated sensors (e.g., piezoelectric sensors in the pavement). The noise signals created by non-vehicle vibrations (e.g., environmental changes) may be falsely recognized as the vehicle-induced vibration signals or overwhelm the vehicle signals, resulting in wrong or missing detection of vehicles. These large and complex noise conditions can further affect the accuracy of vehicle tracking and characterization.
\end{itemize}

\emph{TelecomTM} addresses the above two challenges through a System Characterization module and a Bayesian Analysis using Distributed Sensors module.  In the first module, we characterize the system through driving tests that use a car with a GPS antenna to drive across the road. We estimate the geographic position of each virtual sensor by matching the vehicle's GPS signals with the induced telecom fiber responses. Virtual sensor properties, including signal patterns and transmissibility, are also learned from the driving tests to help design the vehicle detection method and determine its model parameters. In the second module, the arrival times of vehicles at each virtual sensor channel are first estimated through a prominence-based peak detection method. Then, a spatial-domain Bayesian filtering and smoothing algorithm is developed to address the challenge of large and complex noise conditions. It estimates the posterior probability of vehicle arrival time recursively over the space (in the direction of vehicle motion) through fusing spatial-dependent vehicle detection results across multiple virtual sensors. It uses the spatial dependency of distributed sensors and Newton's laws of motion to combine the distributed sensor data to reduce vehicle detection uncertainties and estimate vehicle motion states (positions and moving speed). Vehicle tracking is achieved by converting the estimated arrival times and their derivatives into vehicle positions and speeds. Furthermore, the time differences between responses induced by vehicle wheels and the magnitude of vehicle-induced quasi-static strain are calculated for estimating wheelbase length and vehicle weight, respectively.

We evaluated \emph{TelecomTM} through comprehensive field experiments on an approximately 900-meter road with regular traffic. \emph{TelecomTM} achieves a 90.18\% two-way traffic detection accuracy, 27$\times$ and 5$\times$ error rate reductions for vehicle position tracking and speed tracking, respectively, compared to a baseline method without geo-localization, $\pm$3.92\% percent error (95\% confidence interval) for wheelbase estimation, and $\pm$11.98\% percent error (95\% confidence interval) for weight estimation.

The main contributions of this work are:
\begin{itemize}
    \item We introduce the \emph{TelecomTM} system that uses pre-existing telecom fiber responses induced by vehicle vibrations to enable fine-grained and ubiquitous detection, tracking, and characterization of each vehicle. 
    \item We analyze the telecom fiber's dynamic strain responses to address the challenges of high uncertainties in sensor properties and complex noise conditions through a Bayesian analysis approach. Our approach efficiently characterizes the system and integrates the spatial dependency of distributed sensors to improve vehicle detection, tracking, and characterization accuracy.
    \item We evaluate our system through real-world experiments and characterize the system's performance with various traffic conditions, vehicle types, traveling directions, and speeds. 
\end{itemize}

The rest of the paper is organized as follows: Section 2 describes the physical foundations enabling \emph{TelecomTM}. We introduce our system in Section 3. Our real-world evaluation, its results, and the characterization of our system's performance are described in Section 4. Section 5 discusses related work and the differences between our work and previous research. In Section 6, we conclude our work.

\section{Physical foundations of TelecomTM}

To provide a background understanding of the \emph{TelecomTM} system, we begin with describing its physical foundations, including the principles of distributed acoustic sensing (DAS) and an exhibition of vehicle-induced telecom fiber vibration.

\subsection{Principles of Distributed Acoustic Sensing}

\begin{figure}[!tb]
    \centering
    \includegraphics[width=0.4\linewidth]{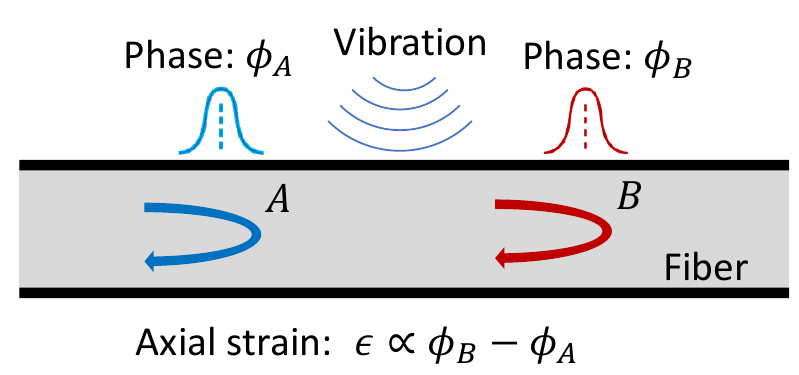}
    \caption{Illustration of the principle of DAS. A strain perturbation affecting the optical fiber caused by vibrations between $A$ and $B$ produces linearly proportional variations in the phase of the backscattered light.}
    \label{fig:das}
\end{figure}

DAS based on the $\phi$-OTDR technique~\cite{shatalin1998interferometric} that uses a standard fiber cable as virtual axial strain sensors. Specifically, an optoelectronic instrument called the interrogator unit repeatedly injects a laser pulse into a fiber cable. An optical interferometry system measures the Rayleigh-backscattered light. The arrival time of the segmented backscattered light can be mapped to the distance along the fiber because the speed of light in the fiber is known. A strain perturbation in the fiber's surroundings may cause a phase shift of the scattering centers. The phase shift is quasi-linearly proportional to the total strain along fiber~\cite{grattan2000optical}. Therefore, the strain variations in different fiber sections can be obtained by repeatedly measuring the phase shift. Figure~\ref{fig:das} illustrates the principle of DAS. By sending the laser pulses at a high frequency (e.g., 250 Hz for \emph{TelecomTM}), the dynamic strain profile along the fiber can be determined~\cite{lellouch2019seismic}. 

Besides the sampling rate, there are two important specifications of DAS systems: gauge length and channel spacing. Gauge length is the length over which the phase shifts are measured. Channel spacing is the distance between each virtual sensor. Selecting the gauge length and channel spacing is a trade-off: a longer gauge length has a higher signal-to-noise ratio but a lower spatial resolution and vice versa. A finer channel spacing can improve the spatial resolution but would have more overlapping sensing areas (i.e., requiring larger data storage) for virtual sensors if the channel spacing is smaller than the gauge length. We chose a 10-meter gauge length and a 1-meter channel spacing in our system to obtain fine-grained traffic information with an adequate signal-to-noise ratio for accurate vehicle detection. In other words, we convert the telecom fiber cable into virtual sensors spatially distributed every meter to measure the strain over each 10-meter section of the fiber.

\subsection{Vehicle-Induced Telecom Fiber Vibration}

\begin{figure}[!tb]
    \centering
    \begin{subfigure}{0.3\textwidth}
        \centering
        \includegraphics[width=1\linewidth]{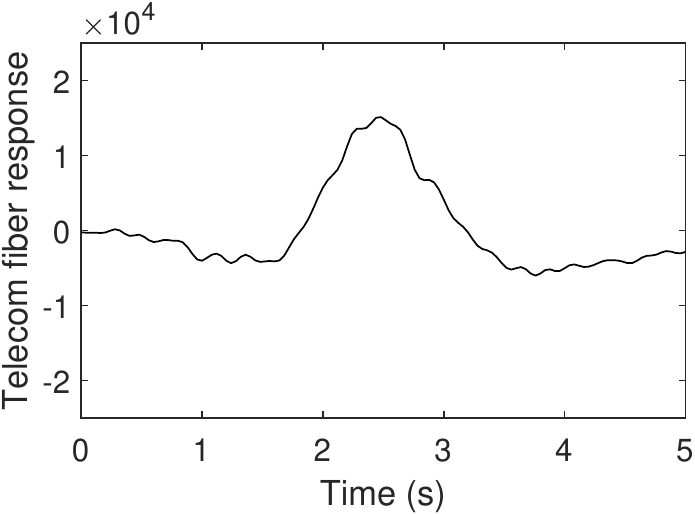}
        \subcaption{}
    \end{subfigure}\qquad
    \begin{subfigure}{0.3\textwidth}
        \centering
        \includegraphics[width=1\linewidth]{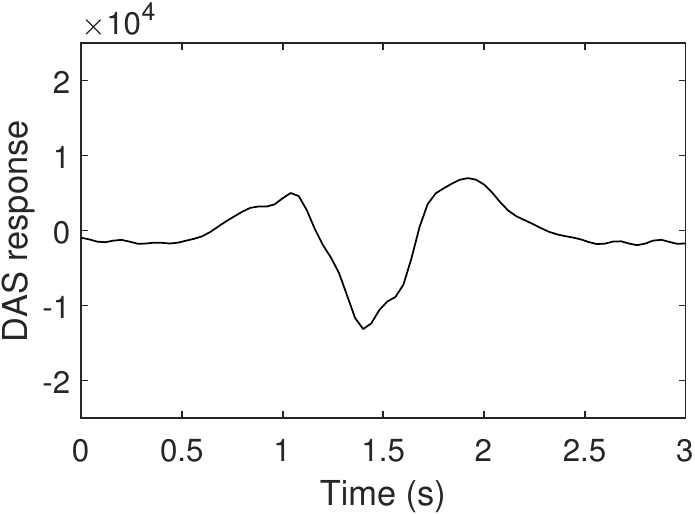}
        \subcaption{}
    \end{subfigure}
    \vspace{-1em}
    \caption{Vehicle-induced telecom fiber response in two virtual sensors having (a) bell-shaped response and (b) polarity-flipped response.}
    \label{fig:responses}
\end{figure}

When a vehicle passes over virtual sensors of the roadside telecom fiber cable, the interaction between the vehicle and the road structure induces the telecom fiber cable to vibrate. The signal pattern of vehicle-induced telecom fiber vibrations depends on the vehicle characteristics, fiber conduit properties, the cable surrounding conditions, etc. There are mainly two components of signals produced by moving vehicles that are recorded by roadside distributed acoustic sensors: 1) quasi-static signals ($<$ 1 Hz) resulting from the ground deformation due to the vehicle's weight, and 2) surface waves (3 to 20 Hz) caused by the dynamic vehicle-road interaction due to the roughness of the road (e.g., bumps). Note that vehicle-induced surface-waves are usually the strongest between 3$\sim$20 Hz. From our observation, vehicles don't excite evident 1$\sim$3 Hz energy, which is relatively weak unless other sources (e.g., earthquakes) exist. Previous studies~\cite{Lindsey2020,yuan2020near} have found that the quasi-static component dominates the energy of vehicle-induced telecom fiber vibration and is theoretically described by the Flamant-Boussinesq approximation \cite{Fung,Ende2021}. As a vehicle approaches the virtual sensor, ground deformation above the sensor increases, and the fiber coupled to the earth is stretched, resulting in increased tension in the fiber. As the vehicle moves away, ground deformation near the virtual sensor and the fiber tension decreases. As a result, the vehicle motion creates a bell-shaped response when it passes a virtual sensor. Figure~\ref{fig:responses} (a) shows an example of the bell-shaped response of a virtual sensor to a passing car that matches the theoretical telecom response. The signal peaks at around 2.5 second when the car reaches the sensor. Positive amplitude indicates the fiber beneath the car is under tension. Since the quasi-static signal created by a moving vehicle dominates the signal energy and is easy to recognize, traffic volume and speed estimation approaches have been developed by detecting, extracting, and localizing these quasi-static signals with the surface-wave component filtered out~\cite{Shen2021,VandenEnde2022a,Ende2021,Lindsey2020,Yuan2021}.

\begin{figure}[!tb]
    \centering
    \includegraphics[width=0.32\linewidth]{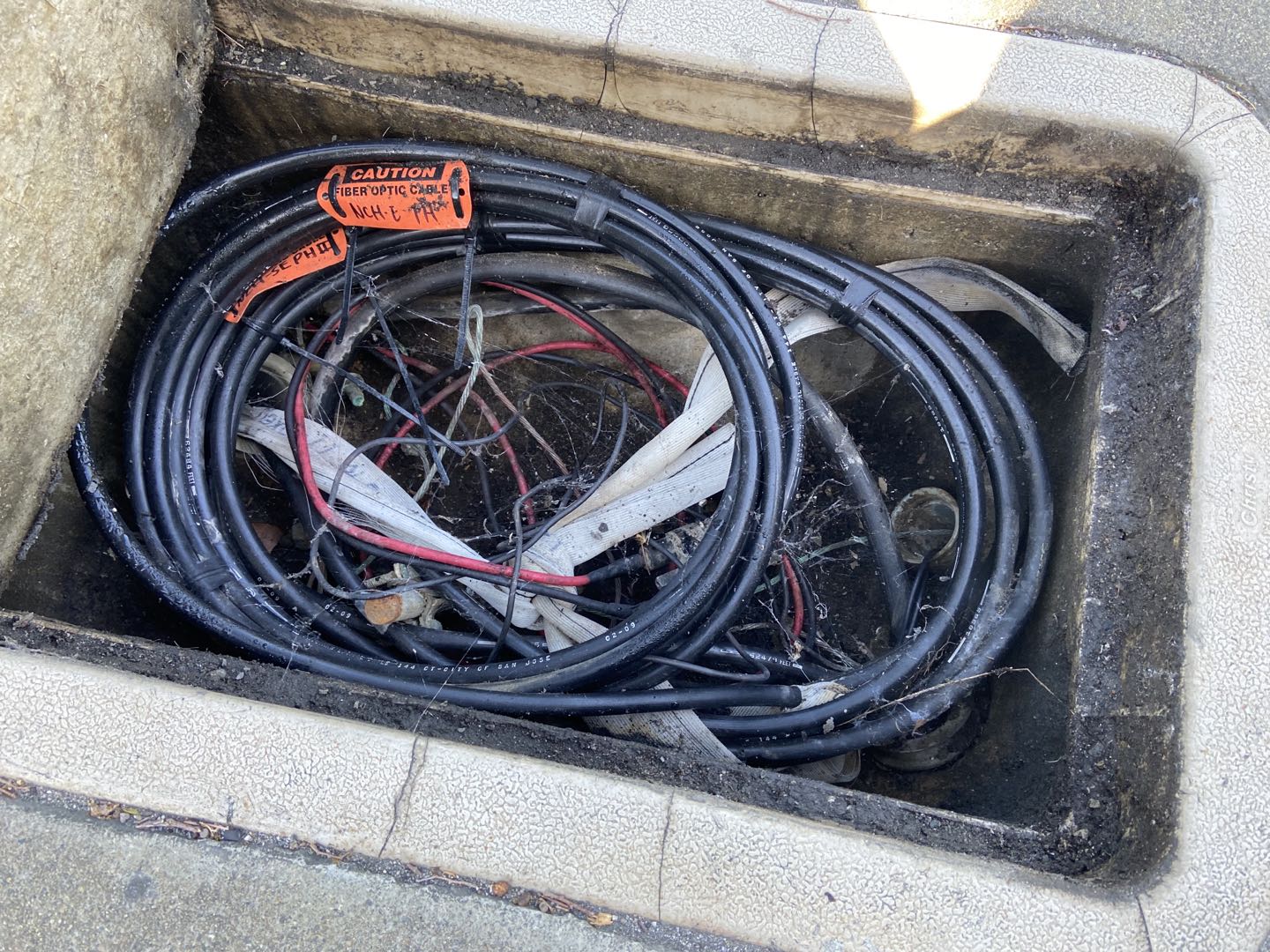}
    \caption{An example photo showing fiber spooling.}
    \label{fig:spooling}
\end{figure}

\begin{figure}[!tb]
    \centering
    \begin{subfigure}{0.32\textwidth}
        \centering
        \includegraphics[width=1\linewidth]{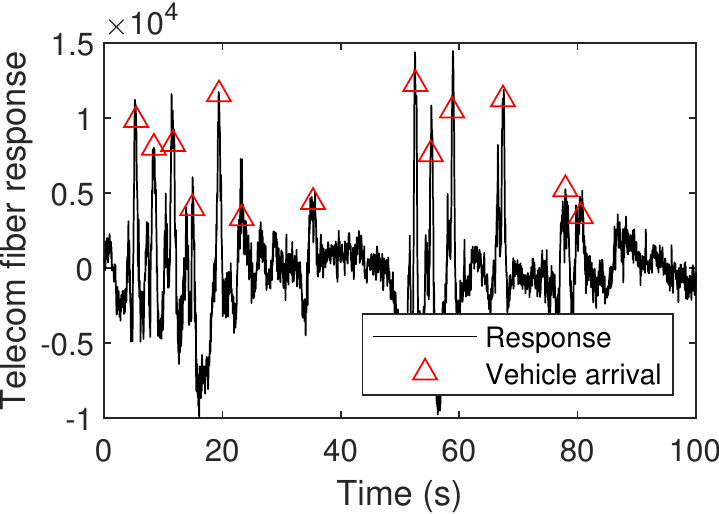}
        \subcaption{}
    \end{subfigure}\qquad
    \begin{subfigure}{0.32\textwidth}
        \centering
        \includegraphics[width=1\linewidth]{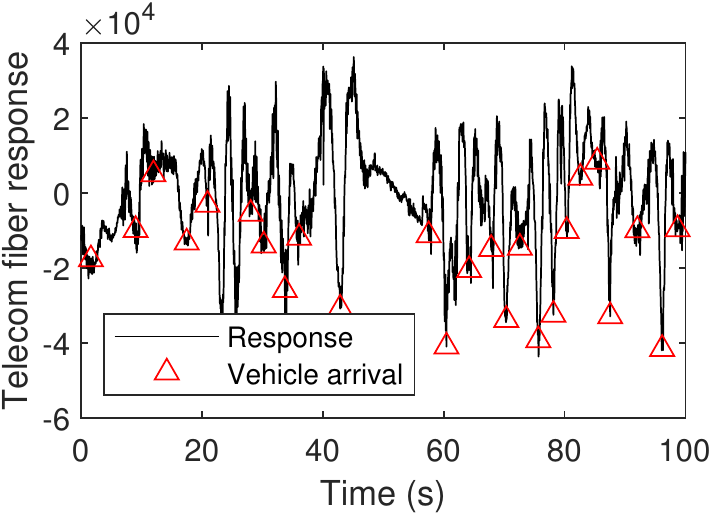}
        \subcaption{}
    \end{subfigure}
    \vspace{-1em}
    \caption{Telecom fiber signal examples of (a) bell-shaped and (b) polarity-flipped response. Red triangle markers indicate arrival times of vehicles. Vehicles are detected at local maximums for the bell-shaped response and at local minimums for the polarity-flipped response.}
    \label{fig:detection}
\end{figure}

However, we observe that due to the unknown sensor properties and complex noise condition challenges mentioned in Section 1, field data do not always match the theoretical vehicle-induced telecom fiber response. First, moving vehicle locations estimated using quasi-static signals do not match the designed locations of virtual sensors based on the maps/drawings of urban fiber installations. It is because these maps are often inaccurate and do not take into account the slack in the fiber cable accumulated underground (e.g., spools in cabinets and manholes). Figure~\ref{fig:spooling} shows an example photo of fiber spooling in a manhole. There are around 100 meters of fibers spooled in this manhole, resulting in a large estimation error of vehicle positions. We also observe from field data that a polarity-flipped response (Figure~\ref{fig:responses} b) consistently occurs at some fiber subsections. The polarity flipping phenomenon could be explained by the near-surface heterogeneity and/or fiber conduit property changes that could lead to stress concentration reversing the fiber's response from tension to compression. In addition, telecom fiber responses have high-frequency and irregular signal backgrounds/trends due to noise and non-vehicle-induced perturbations (e.g., environmental changes) in the surroundings of the telecom fiber cable. Figure~\ref{fig:detection} shows two telecom fiber signals having (a) bell-shaped and (b) polarity-flipped responses. Red triangle markers indicate the arrival of vehicles. We can observe that because of the polarity flipping and irregular signal background phenomenons, conventional peak detection methods, which find local maxima or minima by comparison of neighboring values, or baseline subtraction methods~\cite{elhabian2008moving}, which remove low-frequency or harmonic signal background, may not work well for detecting and tracking vehicles. 

To this end, we introduce our \emph{TelecomTM} system in the next section, which geo-localizes every virtual sensor and analyzes their signal patterns (bell-shaped or polarity-flipped response) by matching the vehicle's position signal with the corresponding vehicle-induced quasi-static signal in telecom fiber. \emph{TelecomTM} overcomes the challenge of large and complex noise conditions by fusing spatially dependent vehicle detection results across multiple virtual sensors. It uses Newton's laws and the spatial dependency of vehicle motion to reduce vehicle detection uncertainty and estimate vehicle positions and speeds. 

\begin{figure}[!tb]
    \centering
    \includegraphics[width=0.7\linewidth]{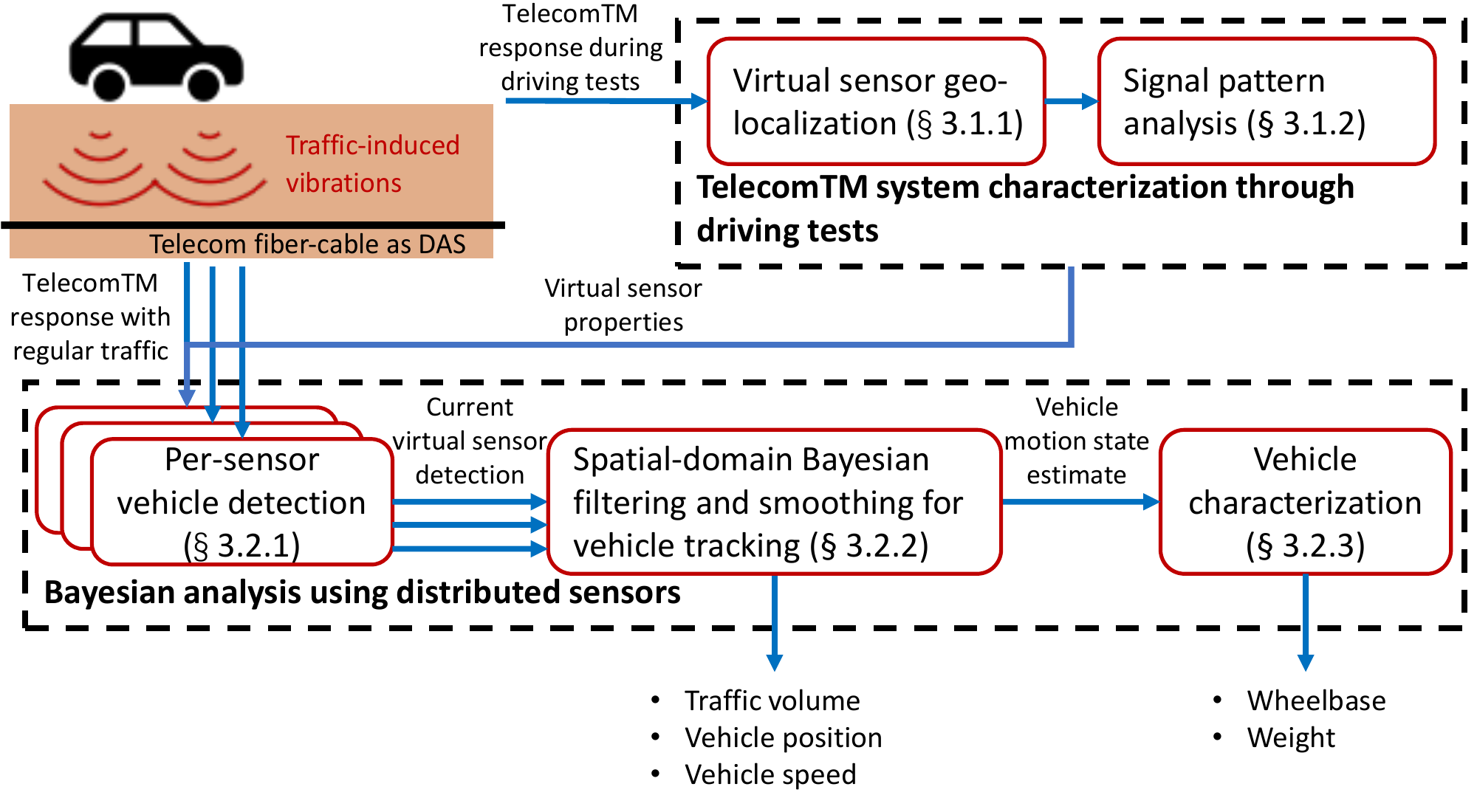}
    \caption{\emph{TelecomTM} system overview.}
    \label{fig:system}
\end{figure}

\section{TelecomTM: Traffic Monitoring System Using Pre-existing Roadside Telecom Fiber cables}

The main goal of \emph{TelecomTM} is to develop a fine-grained and ubiquitous traffic monitoring system that (1) detects and tracks vehicle motion and (2) characterizes vehicles (e.g., vehicle position, speed, wheelbase, and weight) from telecom fiber responses.

\emph{TelecomTM} consists of two modules, as shown in Figure~\ref{fig:system}. The \emph{TelecomTM} System Characterization Module (Section 3.1) conducts driving tests to estimate the geographic position and analyze the signal pattern of each virtual sensor by matching the testing vehicle's GPS recording with the vehicle-induced quasi-static signal in the telecom fiber. In the second module (Section 3.2), telecom fiber responses are inputted to detect vehicles at each virtual sensor using a prominence-based detection method whose parameters are determined using the virtual sensor properties. Then, our system integrates vehicle detection results from spatially-distributed sensors through Bayesian analysis. We estimate the posterior probability of vehicle arrival time using a spatial-domain Bayesian filtering and smoothing algorithm.
Our method combines multiple sensors' information to reduce vehicle detection uncertainties and estimate vehicle motion states (positions and speeds) based on the spatial dependencies of the virtual sensors and Newton's laws of motion. Finally, vehicle characteristics, including weight and wheelbase, are estimated using the vehicle-induced quasi-static signals and the time difference between vehicle wheel-induced responses.

\subsection{TelecomTM System Characterization}

The first module of our system is to characterize each virtual sensor to estimate their geographic locations and analyze their signal patterns.

\subsubsection{Virtual Sensor Geographic Localization}

Accurate mapping of the virtual sensors to their geographic locations is essential for all traffic monitoring applications discussed in the paper. We address the  unknown geo-location challenge by analyzing the telecom fiber response generated by a car with onboard GPS, which is an easily scalable tool when the fiber cables are installed in proximity of public streets~\cite{Yuan2021} (i.e., roadside). In particular, we first conducted ``tap'' tests in the interrogator room (underground and without GPS signal) by hitting the fiber for temporal synchronization between the \emph{TelecomTM} system and the onboard GPS receiver. 
{We then drove the testing car on the road under regular traffic.} The most predominant quasi-static signal peaks when the car is the closest to a virtual sensor. Based on the peaking time and the onboard GPS recording, we can retrieve the geographic positions of every roadside virtual sensor. Also, to reduce GPS error, we run the driving test several times and average the estimated virtual sensors' geo-locations of all the tests. {Note that this characterization process can be done with any city maintenance vehicles, public transit vehicles, mail delivery trucks, etc., which need to go around the city every day. Therefore, it is time- and cost-effective and does not have any effect on regular traffic and telecommunication signals}.

\subsubsection{Vehicle-induced Telecom Fiber Signal Pattern Analysis}
\label{sec:212}

\begin{figure}[!tb]
    \centering
    \begin{subfigure}{0.32\textwidth}
        \centering
        \includegraphics[width=1\linewidth]{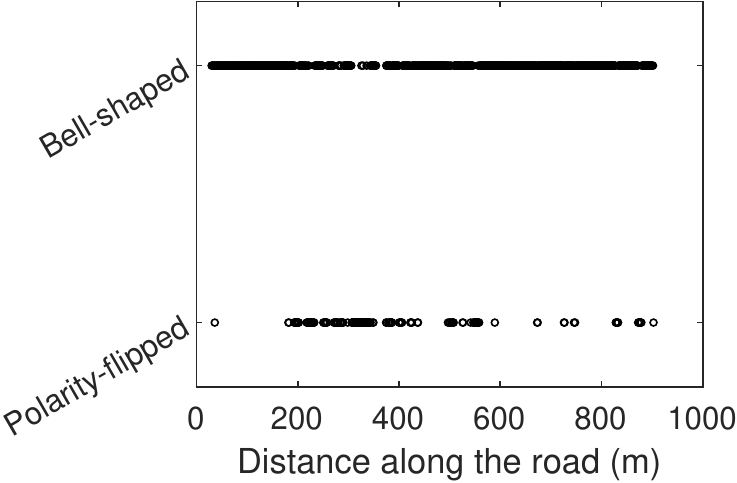}
        \subcaption{}
    \end{subfigure}\qquad
    \begin{subfigure}{0.32\textwidth}
        \centering
        \includegraphics[width=1\linewidth]{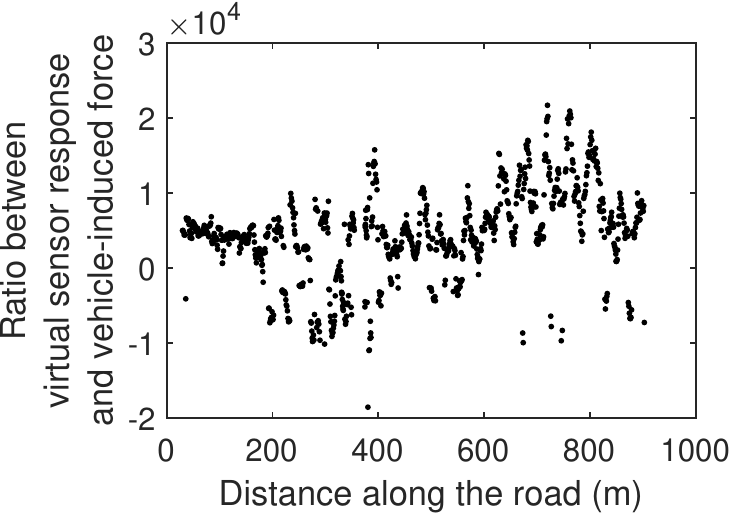}
        \subcaption{}
    \end{subfigure}
    \vspace{-1em}
    \caption{Spatial-variations of telecom fiber signal characteristics: (a) signal pattern and (b) transmissibility (ratio between virtual sensor response and vehicle's weight). Signal properties of distributed virtual sensors have large spatial variations.}
    \label{fig:spatial}
\end{figure}

As mentioned in Section 2.2, telecom fiber responses have various signal patterns, which require different methods and model parameters for detecting vehicles at different virtual sensors. For the bell-shaped response, vehicle motion creates prominence of a peak and should be detected at local maximums; for the polarity-flipped response, vehicle motion creates prominence of a valley and should be detected at local minimums. Also, since the ratio between the virtual sensor's response and vehicle-induced force (i.e., transmissibility) is unknown, a universal vehicle detection method with the same parameters (e.g., peak detection threshold) would fail to detect vehicles accurately. The signal pattern and transmissibility are randomly varying along the fiber, and their distributions are difficult to predict. For example, Figure~\ref{fig:spatial} shows the spatially varied signal pattern and transmissibility along an around 900-meter long road section, having 1120 virtual sensors. 923 of them have bell-shaped responses; 197 of them have polarity-flipped responses. The absolute value of transmissibility varies between 264 and 21704 of telecom fiber response magnitude per ton. To address this heterogeneous and unknown signal characteristics challenge, during our driving tests, we also estimate the transmissibility of each virtual sensor. We define the transmissibility, $T_k$, of the $k$-th sensor as the ratio between the prominence amplitude of the quasi-static signal and the testing vehicle weight. For bell-shaped responses, $T_k>0$, and for polarity-flipped responses, $T_k<0$. {In addition, by using the signal of our testing vehicle passing over the fiber on the nearest lane and the physical model that describes the distribution of subsurface stresses (Boussinesq's theory), we can estimate the signal transmissibilities at further lanes without the need for additional driving tests.}

\subsection{Bayesian Analysis Using Distributed Sensors}
The second module of \emph{TelecomTM} is a three-step Bayesian analysis approach. The per-sensor detection step estimates the arrival times of vehicles at each virtual sensor by detecting the prominence of peaks or valleys in time-domain signals. The second step integrates spatially dependent information across various adjacent virtual sensors to reduce false positive detections and estimate vehicle motion states (positions and speeds). The third step estimates the wheelbase lengths and weights of the tracked vehicles.

\subsubsection{Per-sensor Vehicle Detection}
In this step, we detect vehicles by estimating vehicles' arrival times at each virtual sensor from the time-domain telecom fiber responses. As we discussed in Section 2.2, the quasi-static component due to the vehicle's weight dominates the energy of the signal. Therefore, we use a prominence-based detection method to detect vehicle occurrence. The prominence of a peak measures how much the peak stands out due to its intrinsic height and its relative location to other peaks. We use the prominence-based detection method because it is more robust for detecting local minima or maxima of a signal having high-frequency and irregular signal background. Specifically, we first smooth the data using the locally weighted smoothing (LOESS)~\cite{cleveland1988locally} with a smoothing span of a one-second window. Here we remove outliers and high-frequency noise that could be incorrectly detected as prominence. In addition, our per-sensor detection method determines the prominence detection threshold of each sensor using its transmissibility. The larger the transmissibility, the larger the prominence threshold. It allows our method to be adaptive to sensors with heterogeneous properties. Specifically, for sensors with $T_k>0$, the prominence of a peak larger than $r_0\times (|T_k|/T_0)$ within a one-second window is detected as a vehicle, and for sensors with $T_k\leq 0$, prominence of a valley is detected. $T_0$ is the minimum absolute transmissibility ($T_0=\min(|T_k|)$, for $k=1,\cdots, K$). $K$ is the total number of virtual sensors. $r_0$ is the prominence threshold for vehicle detection at the sensor with the transmissibility of $T_0$. 

\subsubsection{Spatial-Domain Bayesian Filtering and Smoothing for Vehicle Tracking}

Although an efficient system characterization is conducted through the driving tests mentioned in the first module of \emph{TelecomTM}, the large and complex noise conditions challenge would still cause false or missing detection of vehicles using only the per-sensor detection method. For instance, the noise signals may be incorrectly recognized as vehicle-induced telecom fiber signals or overwhelm the vehicle signals. Wrong or missing vehicle detection results can further affect the accuracy of vehicle tracking and characterization.

To overcome the challenge, we develop a spatial-domain Bayesian filtering and smoothing algorithm that fuses spatial-dependent vehicle detection results of distributed virtual sensors to improve vehicle detection and tracking accuracy. Since vehicle motion is continuous in the spatial domain, our algorithm estimates vehicle motion states based on Newton's laws of motion. 
In particular, we first formulate our vehicle detection and tracking problem as a spatial-domain state-space model. State-space model is a more compact and convenient representation for multiple-input multiple-output dynamic systems compared to other representations, such as the transfer function model and impulse response function model~\cite{keesman2011system}. In addition, to use the spatial dependency of distributed virtual sensors, our formulation considers the arrival time of a vehicle at every virtual sensor as measurement. It is different from the conventional formulation of Bayesian filtering and smoothing for motion tracking or navigation~\cite{rauch1965maximum,sarkka2013bayesian,zarchan2000fundamentals} whose measurement/observation is the position of the object at every timestep and does not take into account the spatial dependency information. The vehicle arrival time at the $k$-th virtual sensor, $t$, and its derivative, $\dot{t}$, are described by the linear state space ${\bf t}_k = [t,\dot{t}]^T$. The derivative of arrival time, $\dot{t}$, is the time for the vehicle to travel one meter. Based on Newton's laws of motion, the state-space model is
\begin{equation}
\begin{aligned}
    & {\bf t}_k={\bf A}_k{\bf t}_{k-1}+{\bf w}_k\\
    & {z}_k={\bf C}{\bf t}_k+{ v}_k
\end{aligned}
\end{equation}
$${\bf A}_k = \begin{bmatrix}1 & \Delta x_k\\
0 & 1\end{bmatrix},~{\bf C}=[1,~0],~{\bf w}_k\sim\mathcal{N}(0,{\bf Q}_k),~{ v}_k\sim\mathcal{N}(0,\sigma_z)$$
$${\bf Q}_k = \begin{bmatrix}\frac{1}{4}\Delta x_k^4 & \frac{1}{2}\Delta x_k^3\\
\frac{1}{2}\Delta x_k^3 & \Delta x_k^2\end{bmatrix}\sigma_{\ddot{t}}^2,$$
where $\Delta x_k$ is the distance between the $(k-1)$-th and the $k$-th virtual sensor projected on the road; ${z}_k$ is the observed vehicle arrival time at virtual sensor $k$; ${\bf w}_k$ is the process noise; ${\bf v}_k$ is the observation noise; and $\sigma_{\ddot{t}}$ and $\sigma_z$ are the standard deviations of arrival time's second derivative and that of the measurement noise, respectively.

\begin{figure}[!tb]
    \centering
    \includegraphics[width=0.8\linewidth]{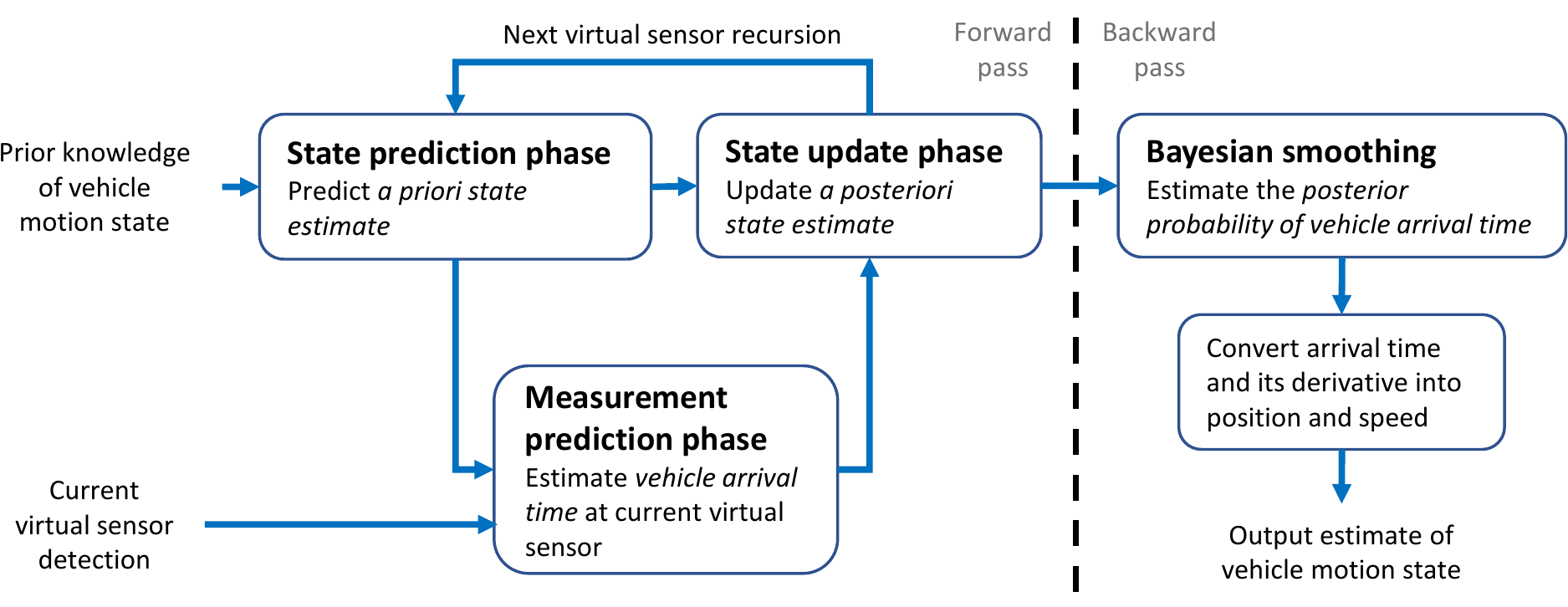}
    \caption{Our Bayesian filtering and smoothing algorithm.}
    \label{fig:bayes}
\end{figure}

To estimate the vehicle motion state (described in the above state-space model) from multiple sensors' noisy data, we develop a Bayesian filtering and smoothing algorithm (as shown in Figure~\ref{fig:bayes} and Algorithm~\ref{alg:bayes}). It consists of a forward filtering pass and a backward smoothing pass. The forward pass estimates the filtered posterior probability of the state estimate recursively using past observations; the backward pass computes the smoothed posterior probability using all observations~\cite{sarkka2013bayesian,bryson2018applied}. Specifically, the forward pass of our algorithm estimates the posterior probability of vehicle arrival time over space using vehicle detection results of previous and current virtual sensors: $p({\bf t}_k|{z}_{1:k})$. The forward pass has three phases: 1) state prediction phase, 2) measurement prediction phase, and 3) state update phase. The state prediction phase uses the state estimate from the previous virtual sensors to produce the state estimate at the current virtual sensor. Since we don't have direct measurement/observation (i.e., the arrival time of the vehicle) at each virtual sensor, our algorithm has a measurement prediction phase that predicts the measurement, ${z}_k$, by finding the detection in the current virtual sensor having the largest probability of the predicted arrival time:
\begin{equation}
\label{eq:bayes}
    \hat{z}_k=\arg\max_{z\in {\bf D}_k} f(z;\hat{t}_{k|k-1}, \sigma_{t,k|k-1})
\end{equation}
where ${\bf D}_k$ are per-sensor vehicle detection results for the virtual sensor $k$; $f(z)$ is the probability density function of the predicted arrival time following the Gaussian distribution: $\mathcal{N}(z;\hat{{t}}_{k|k-1},\sigma_{t,k|k-1}^2)$; $\hat{t}_{k|k-1}$ and $\sigma_{t,k|k-1}$ are the predicted arrival time estimate and its standard deviation, respectively. In other words, our algorithm tracks the vehicle arriving at each virtual sensor by finding the closest vehicle detection to the predicted arrival time estimate. Then, the state update phase updates the state estimate (i.e., a posteriori state estimate) using a weighted average of the predicted state estimates and the measurement. The backward pass of our algorithm adopts the Rauch-Tung-Striebel smoother~\cite{rauch1965maximum} to estimate the posterior probability of vehicle arrival time using vehicle detection results at all virtual sensors: $p({\bf t}_k|{z}_{1:K})$. Here our algorithm reduces detection uncertainties by utilizing the spatial dependency of the distributed sensors. Finally, vehicle motion state estimation or tracking is achieved by converting arrival time and its derivative at every virtual sensor, $[t,\dot{t}]$, into vehicle position and speed, $[x,v]$ using the virtual sensors' geographic locations estimated in our module one.

{\small\begin{algorithm}[h!]
\caption{Spatial-domain Bayesian filtering and smoothing}
\label{alg:bayes}
\begin{algorithmic}[1]
\INPUT{Vehicle detection results at every virtual sensor, ${\bf D}_k$, for $k=1,\cdots,K$; Prior knowledge of standard deviations, $\sigma_{\ddot{t}}$ and $\sigma_z$}
\OUTPUT{Posterior probability of vehicle state, $p({\bf t}_k|z_{1:K})$}
\STATE \#\emph{Forward filtering pass:}
\STATE {\bf Initialization}: initial guess of state estimate and estimate covariance, ${\bf t}_{0|0}$ and ${\bf P}_{0|0}$
\FOR{Virtual sensor $k=1,\cdots,K$}
    \STATE \#\emph{State prediction phase:}
    \STATE{Calculate the predicted state estimate, $$\hat{\bf t}_{k|k-1}={\bf A}_k\hat{\bf t}_{k-1|k-1}$$}
    \STATE{Calculate the predicated state covariance, $${\bf P}_{k|k-1}={\bf A}_k{\bf P}_{k-1|k-1}{\bf A}_k^T+{\bf Q}_k$$}
    \STATE{Predict the state probability given previous measurement, $$p({\bf t}_{k}|{ z}_{1:k-1})=\mathcal{N}({\bf t}_{k};\hat{\bf t}_{k|k-1},{\bf P}_{k|k-1})$$}
    \STATE \#\emph{Measurement prediction phase:}
    \STATE{Find the vehicle detection of the $k$-th virtual sensor having the largest probability of the predicted state using Equation~\eqref{eq:bayes}}
    \STATE \#\emph{State update phase:}
    \STATE{Calculate the updated state estimate, \begin{align*}
        \hat{\bf t}_{k|k}=&\hat{\bf t}_{k|k-1}+{\bf P}_{k|k-1}{\bf C}^T\\
        &\times({\bf C}{\bf P}_{k|k-1}{\bf C}^T+\sigma_z)^{-1}({ z}-{\bf C}\hat{\bf t}_{k|k-1})
    \end{align*}}
    \STATE{Calculate the updated state covariance, $${\bf P}_{k|k}=({\bf I}-{\bf P}_{k|k-1}{\bf C}^T({\bf C}{\bf P}_{k|k-1}{\bf C}^T+\sigma_z)^{-1}{\bf C}){\bf P}_{k|k-1}$$}
    \STATE{Update the state probability given previous and current measurement, $$p({\bf t}_k|z_{1:k})=\mathcal{N}({\bf t}_k;\hat{\bf t}_{k|k},{\bf P}_{k|k})$$}
\ENDFOR
\STATE \#\emph{Backward smoothing pass:}
\STATE {\bf Initialization}: Smoothed state estimate and covariance at the last virtual sensor, $\hat{\bf t}_{K|1:K}=\hat{\bf t}_{K|K}$, ${\bf P}_{K|1:K}={\bf P}_{K|K}$
\FOR{Virtual sensor $k=K-1,\cdots,1$}
    \STATE{Calculate the smoothed state estimate, \begin{align*}
        \hat{\bf t}_{k|1:K}=\hat{\bf t}_{k|k}+{\bf P}_{k|k}{\bf A}_{k+1}^T{\bf P}^{-1}_{k+1|k}(\hat{\bf t}_{k+1|1:K}-\hat{\bf t}_{k+1|k})
    \end{align*}}
    \STATE{Calculate the smoothed state covariance, \begin{align*}
        {\bf P}_{k|1:K}=&{\bf P}_{k|k}+{\bf P}_{k|k}{\bf A}_{k+1}^T{\bf P}^{-1}_{k+1|k}\\
        &\times({\bf P}_{k+1|1:K}-{\bf P}_{k+1|k})({\bf P}_{k|k}{\bf A}_{k+1}^T{\bf P}^{-1}_{k+1|k})^T
    \end{align*}}
    \STATE{Estimate the posterior state probability given all sensors' measurement, $$p({\bf t}_k|z_{1:K})=\mathcal{N}({\bf t}_k;\hat{\bf t}_{k|1:K},{\bf P}_{k|1:K})$$}
\ENDFOR
\end{algorithmic}
\end{algorithm}}

\subsubsection{Vehicle Characterization}

\begin{figure}[!tb]
    \centering
    \begin{subfigure}{0.3\textwidth}
        \centering
        \includegraphics[width=1\linewidth]{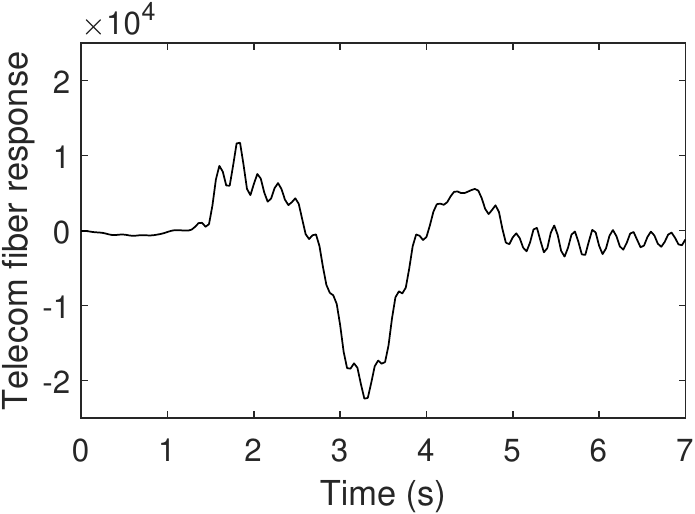}
        \subcaption{}
    \end{subfigure}\qquad
    \begin{subfigure}{0.3\textwidth}
        \centering
        \includegraphics[width=1\linewidth]{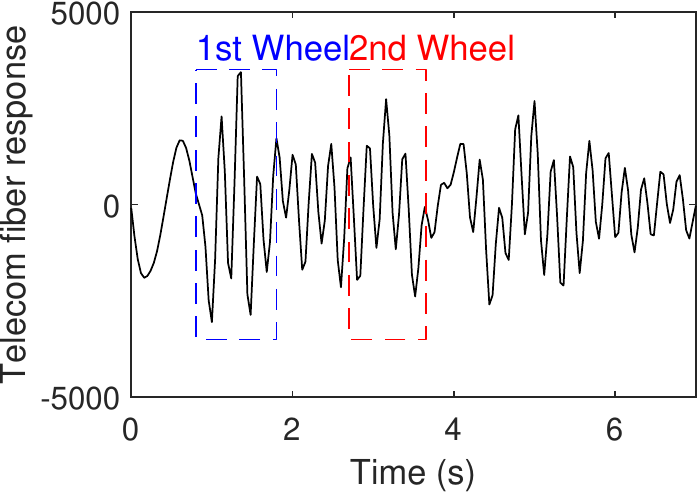}
        \subcaption{}
    \end{subfigure}
    \vspace{-1em}
    \caption{(a) Vehicle-induced telecom fiber responses and (b) its highpass filtered response at a bridge-road joint. Vehicle wheel-induced vibration responses have a repeating impulse response pattern.}
    \label{fig:bridge_rep}
\end{figure}

\emph{TelecomTM} estimates vehicle wheelbase lengths and weights using the vehicle speed estimation results and the quasi-static component of the vehicle-induced signals. To estimate the wheelbase length, we use the high-frequency telecom fiber response ($\geq$3 Hz) created by a vehicle's two wheels passing a bump, joint, or pothole. It creates clear impulse responses due to the interaction between the vehicle wheels and the abrupt change in road profiles. Figure~\ref{fig:bridge_rep} shows an example of a virtual sensor response and its highpass filtered signal ($\geq$3 Hz) at a bridge-road joint. We can observe the vibration components created by the two wheels have a repeating impulse response pattern. The time difference between the two wheel-induced responses can be estimated by using the auto-correlation function to find the repeating patterns. Then, we can estimate the wheelbase length by multiplying the time difference with the vehicle speed, $\tau_m\times v$, where $\tau_m=\arg\max_{\tau}R_{zz}(\tau)$ is the time difference between wheel-induced telecom fiber response; $R_{zz}(\tau)$ is the auto-correlation function at lag $\tau$ of the highpass filtered response of the vehicle-induced telecom fiber vibration; and $v$ is the moving speed of the vehicle. 

\begin{figure}[!tb]
    \centering
    \begin{subfigure}{0.35\textwidth}
        \centering
        \includegraphics[width=1\linewidth]{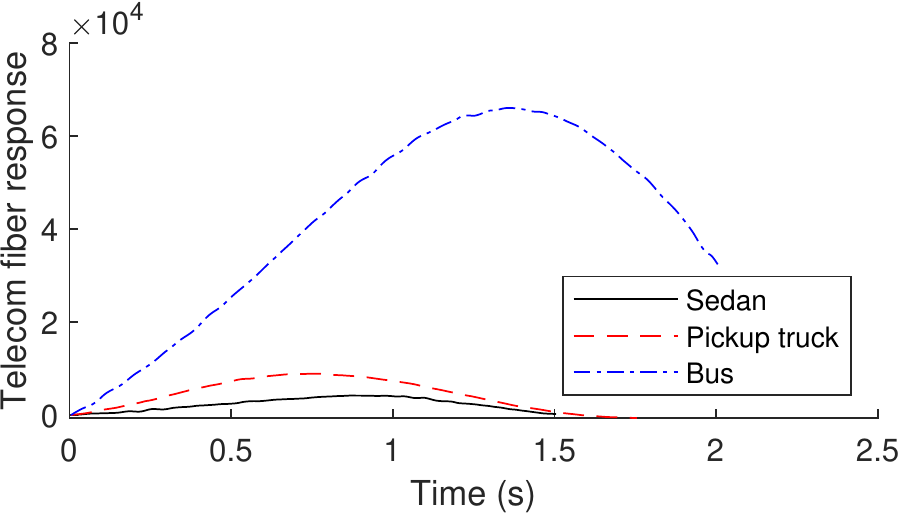}
        \subcaption{}
    \end{subfigure}\qquad
    \begin{subfigure}{0.35\textwidth}
        \centering
        \includegraphics[width=1\linewidth]{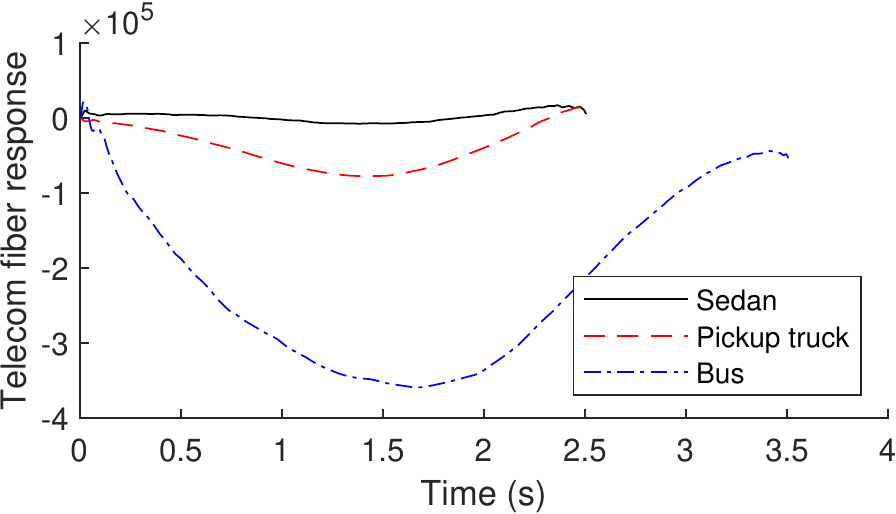}
        \subcaption{}
    \end{subfigure}
    \vspace{-1em}
    \caption{Quasi-static signals of (a) bell-shaped and (b) polarity-flipped responses for three different types of vehicle. Heavier vehicle creates larger prominence amplitude.}
    \label{fig:wim}
\end{figure}

Furthermore, by assuming the linear elasticity of the road and the near-surface structure, the prominence amplitude of the quasi-static signal is approximately proportional to the moving vehicle's weight. Figure~\ref{fig:wim} shows the telecom fiber responses having bell-shaped and polarity-flipped signal patterns created by three different types of vehicles moving in the same lane and measured by the same virtual sensor. We can observe that the heavier the vehicle, the larger the prominence amplitude. Therefore, we estimate the moving vehicle's weight by computing the weighted average of quasi-static signal prominence, $\frac{1}{K}\sum_{k=1}^K (P_k/|T_k|)$, where $P_k$ is the prominence amplitude of the vehicle's quasi-static signal at virtual sensor $k$.

\section{Real-World Evaluation}

In this section, we describe the experimental setup and results evaluating the performance of \emph{TelecomTM}, followed by a characterization of the system's performance. 

\subsection{Experimental Setup}

\begin{figure}[!tb]
    \centering
    \includegraphics[width=0.6\linewidth]{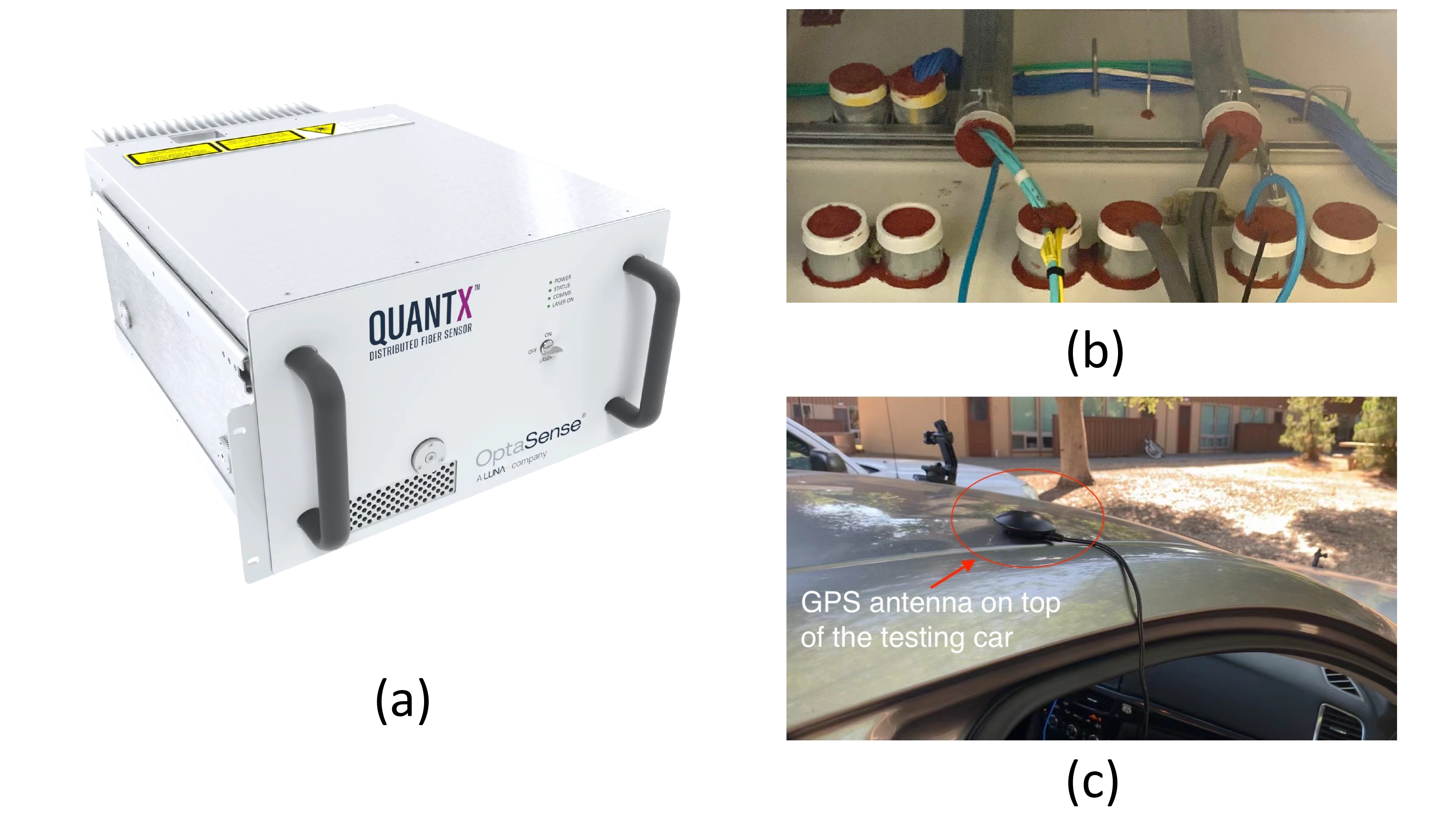}
    \caption{Experimental setup: (a) the QUANTX interrogator~\cite{quantx}, (b) an example photo of fiber conduits, and (c) GPS receiver on the testing car.}
    \label{fig:hardware}
\end{figure}

\begin{figure}[t]
    \centering
    \includegraphics[width=0.5\linewidth]{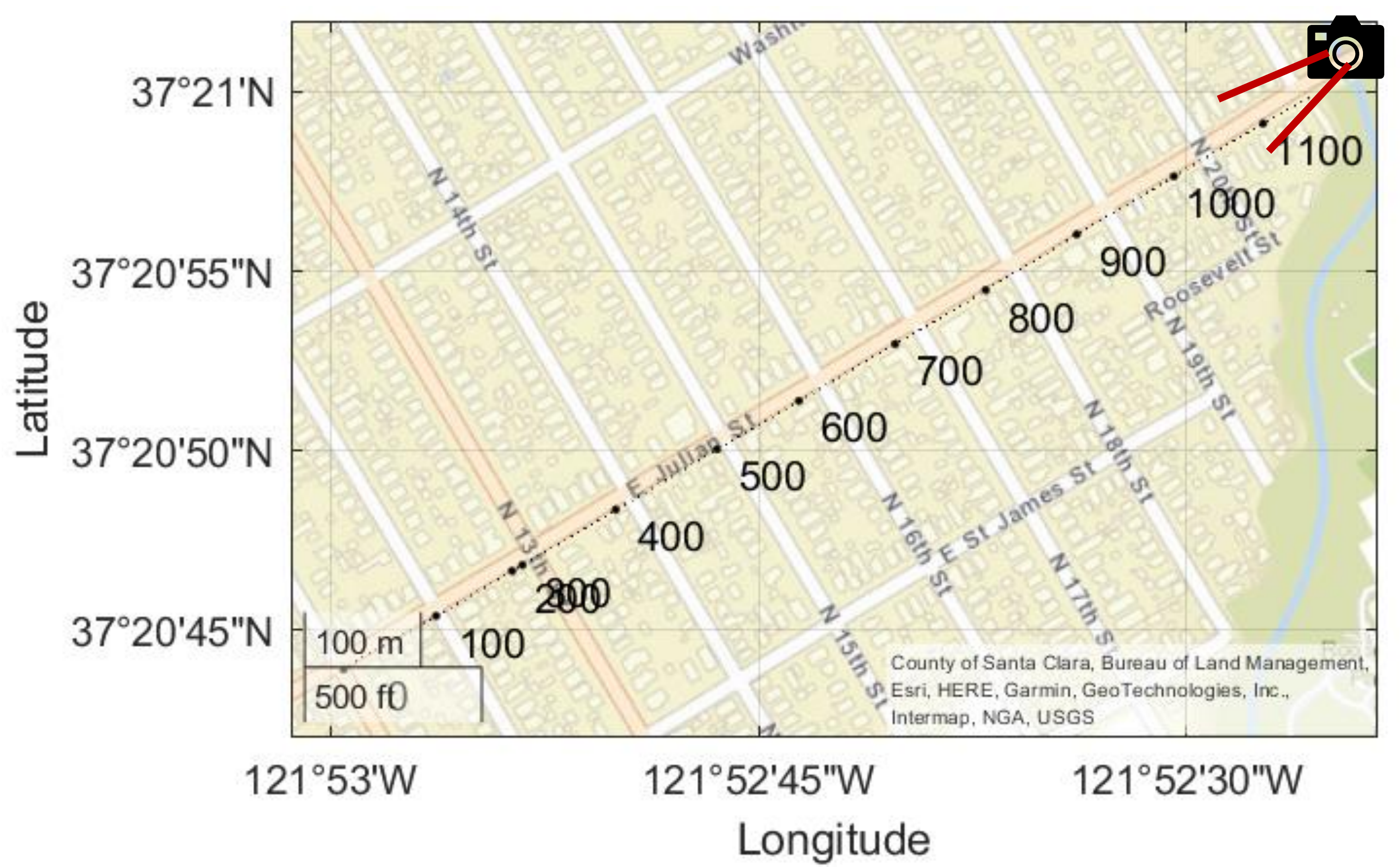}
    \caption{Locations of virtual sensors. The camera icon indicates the location of a camera. The dot line and numbers indicate virtual sensors' locations}
    \label{fig:gps}
\end{figure}

We evaluate \emph{TelecomTM} on a two-way two-lane road with pre-existing telecom fiber cables in San Jose, California. \emph{TelecomTM} responses were collected using an OptaSense QuantX interrogator~\cite{quantx} (as shown in Figure~\ref{fig:hardware} a) at a 250 Hz sampling rate, 10-meter gauge length, and 1-meter virtual sensor spacing. The interrogator recorded dynamic strain responses from a roadside telecom fiber cable (an example photo of fiber conduits is shown in Figure~\ref{fig:hardware} b) along an around 900-meter road section of the East Julian St, San Jose, CA. Figure~\ref{fig:gps} shows the location of the telecom fiber cable that has 1120 virtual sensors. The outbound traffic (with virtual sensor number increasing in Figure~\ref{fig:gps}) travels away from the interrogator, and is closer to the telecom fiber cable than inbound direction traffic.

To estimate geographic locations of virtual sensors and provide ground-truth of vehicle positions, we installed a GPS receiver that records its geographical position every second on a testing sedan vehicle (1.47 tons weight), as shown in Figure~\ref{fig:hardware} (c). We also logged the vehicle speed at a 100 Hz sampling rate through the Controller Area Network (CAN) bus. In addition, a camera was placed close to the last virtual sensor (as shown in Figure~\ref{fig:gps}) to acquire ground-truth information on vehicle arrival times and vehicle models. We obtain wheelbase lengths and weights of vehicles with the vehicle model information. With the supervision of staff from the City of San Jose, we conducted two daytime experiments with 388 vehicles recorded in a 31-minute video and seven nighttime experiments with our testing vehicle running through the testing road.

Our system was evaluated for vehicle detection, tracking, and wheelbase and weight estimations. It achieves overall 90.18\% accuracy for vehicle detection, 27$\times$ error rate reduction (average mean absolute error (MAE) reduced from 140.09 m to 5.16 m) for vehicle position estimation, 5$\times$ error rate reduction (average MAE reduced from 17.18 km/h to 3.57 km/h) for vehicle speed estimation compared to a baseline method {that does not geo-localize the virtual sensors through the driving test.  The baseline method directly applies our spatial-domain Bayesian analysis algorithm introduced in Section 3.2 to the telecom cable responses.} In addition, \emph{TelecomTM} achieves $\pm$3.92\% accuracy for wheelbase estimation, and $\pm$11.98\% percent error for weight estimation. {The performance of \emph{TelecomTM} was found to remain consistent during evaluations conducted on weekdays and weekends, as well as during daytime and nighttime, despite variations in temperature. However, the evaluations were limited to sunny weather conditions only.} We discuss our results in detail in the following subsections.

\subsection{Vehicle Detection Results}

\begin{figure}[!tb]
    \centering
    \begin{subfigure}{0.3\textwidth}
        \centering
        \includegraphics[width=1\linewidth]{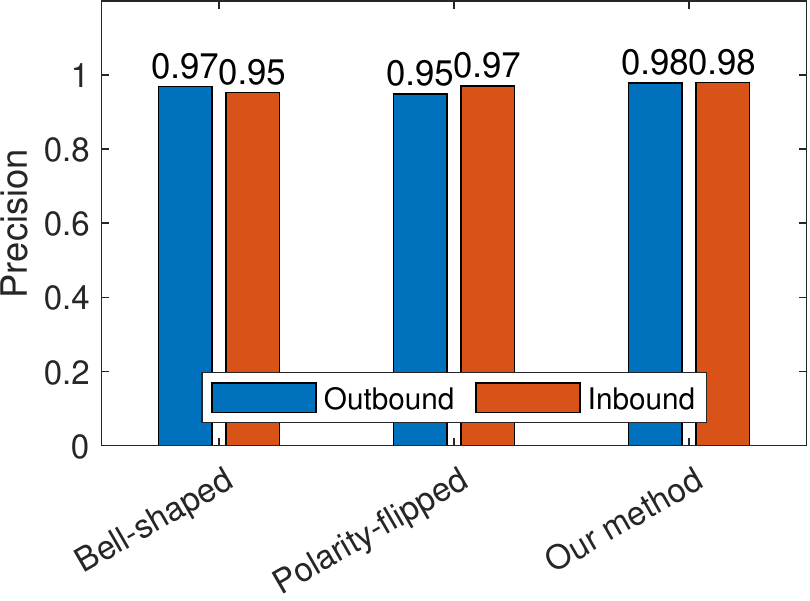}
        \subcaption{}
    \end{subfigure}\qquad
    \begin{subfigure}{0.3\textwidth}
        \centering
        \includegraphics[width=1\linewidth]{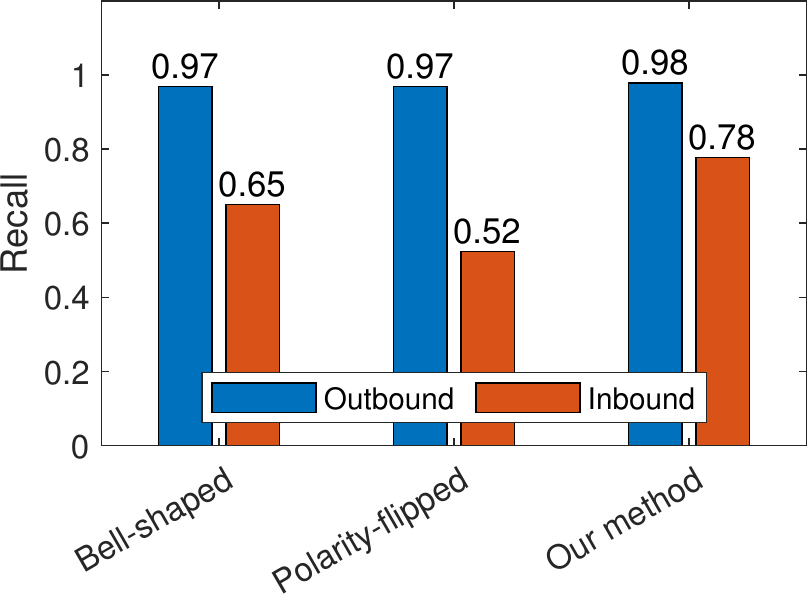}
        \subcaption{}
    \end{subfigure}
    \vspace{-1em}
    \caption{(a) Precision and (b) recall of detection results using only bell-shaped responses, polarity-flipped responses, or using our method with all distributed sensors. Our method improves vehicle detection recall by fusing spatial dependency information from distributed virtual sensors.}
    \label{fig:det_result}
\end{figure}

We compare the vehicle detection results using \emph{TelecomTM} with the ground truths captured using the camera. Figure~\ref{fig:det_result} shows the average precision and recall of detection results using only bell-shaped responses, polarity-flipped responses, or using our \emph{TelecomTM} method. All three cases achieve $\geq 95\%$ precision for both inbound and outbound directions and $\geq 95\%$ recall for the outbound direction. Recall for the inbound direction is lower than that for the outbound direction mainly due to the long distance from the inbound traffic to the telecom fiber cable, which will be discussed in Section 4.5.1. The average recall for vehicle detection using the bell-shaped response is 13\% higher than that using the polarity-flipped response. It is because there are multiple factors causing the polarity-flipping phenomenon, which has high uncertainties, resulting in the distinct signal patterns of the polarity-flipped responses across various virtual sensors. Our \emph{TelecomTM} method has been successful in addressing this issue. It has a 1.6$\times$ improvement in error rate reduction (false negative rate reduced from 35\% to 22\%) by considering spatially dependent vehicle detection information of distributed virtual sensors.

\subsection{Vehicle Tracking Results}

\begin{figure}[!tb]
    \centering
    \begin{subfigure}{0.4\textwidth}
        \centering
        \includegraphics[height=4.5cm]{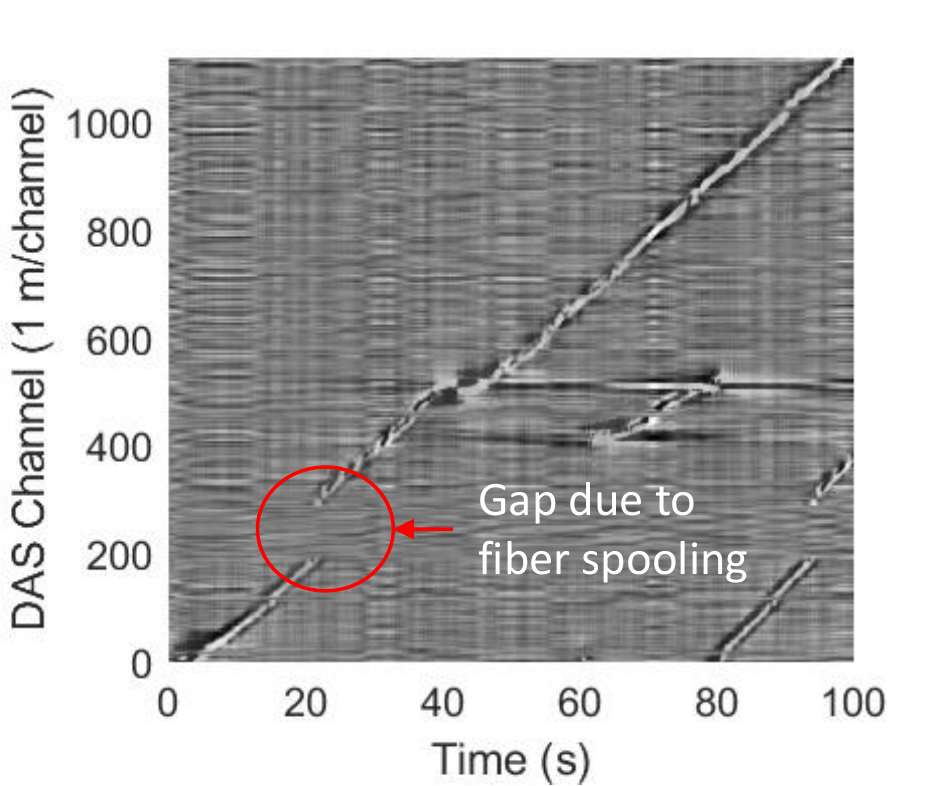}
        \subcaption{}
    \end{subfigure}\qquad
    \begin{subfigure}{0.4\textwidth}
        \centering
        \includegraphics[height=4.5cm]{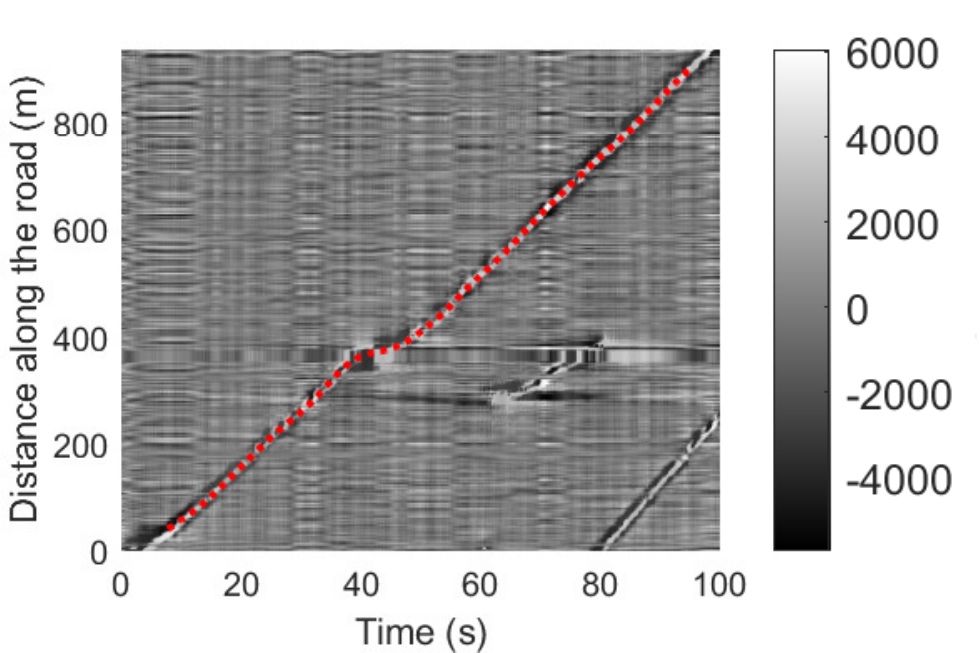}
        \subcaption{}
    \end{subfigure}
    \caption{Telecom fiber response with the y-axis being (a) the fiber distance to virtual sensors or (b) the distance along the road after geo-localization. Red dot curve in (b) shows our vehicle tracking result. Our method removed the signal gap (red circle) due to fiber spooling and estimated the geographic locations of virtual sensors by matching the testing vehicle signal and the corresponding telecom fiber response.}
    \label{fig:single_tracking}
\end{figure}

\begin{figure}[!tb]
    \centering
    \begin{subfigure}{0.3\textwidth}
        \centering
        \includegraphics[width=1\linewidth]{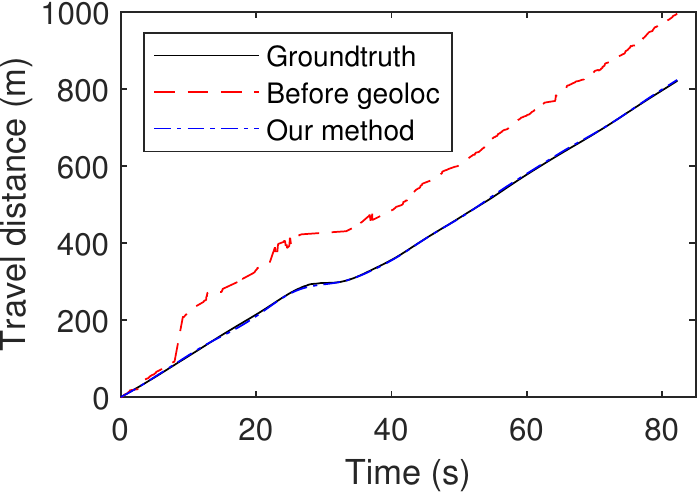}
        \subcaption{}
    \end{subfigure}\qquad
    \begin{subfigure}{0.3\textwidth}
        \centering
        \includegraphics[width=1\linewidth]{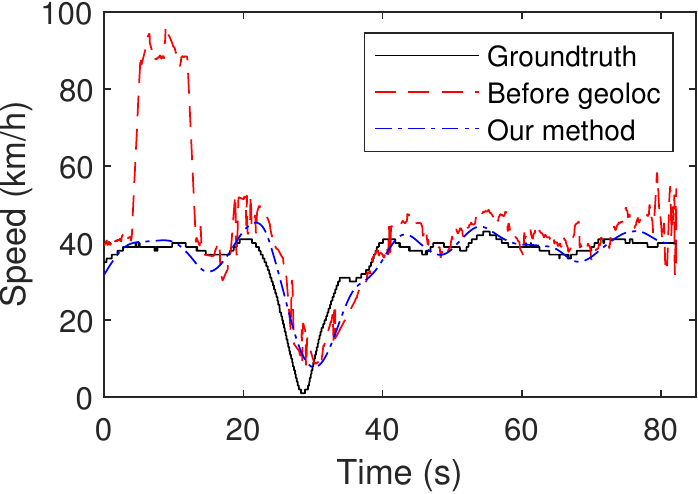}
        \subcaption{}
    \end{subfigure}
    \vspace{-1em}
    \caption{Vehicle (a) locations and (b) speeds ground truth (black curve) and estimations before (blue curve) and after (red curve) geo-localization (geoloc).}
    \label{fig:tracking_result}
\end{figure}

\begin{figure}[!tb]
    \centering
    \begin{subfigure}{0.3\textwidth}
        \centering
        \includegraphics[width=1\linewidth]{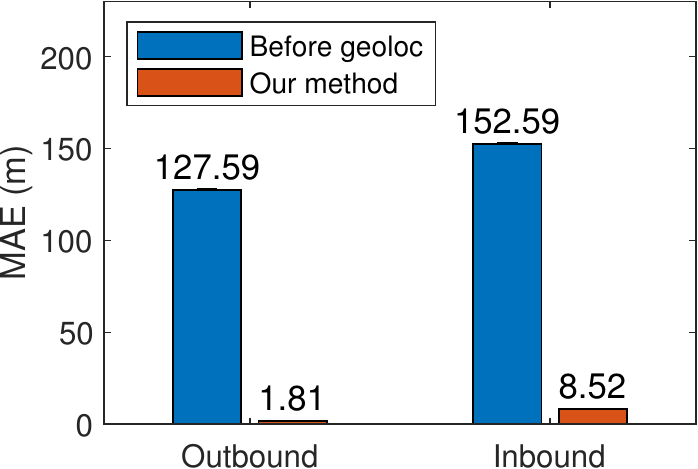}
        \subcaption{}
    \end{subfigure}\qquad
    \begin{subfigure}{0.3\textwidth}
        \centering
        \includegraphics[width=1\linewidth]{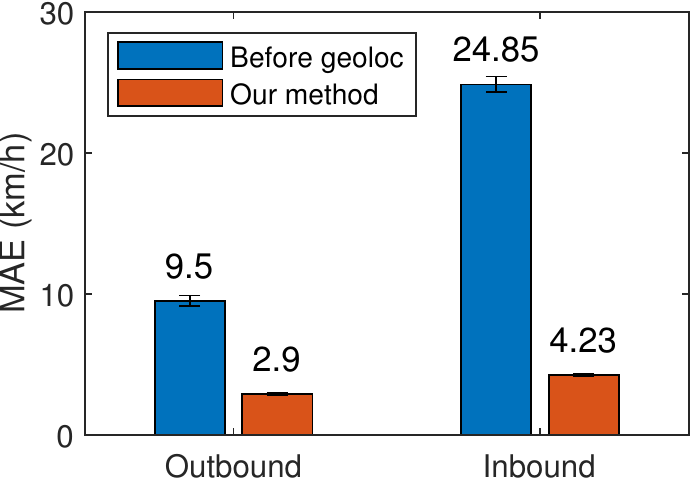}
        \subcaption{}
    \end{subfigure}
    \vspace{-1em}
    \caption{{Mean absolute value bar chart with a 95\% confidence interval} for vehicle (a) location and (b) speed estimations before and after geo-localization (geoloc). {The number above each bar indicates the mean absolute value.} Our method improves the vehicle location and speed estimation accuracy by geo-localizing each virtual sensor.}
    \label{fig:tracking_bar}
\end{figure}

We evaluated the effectiveness of \emph{TelecomTM} for vehicle tracking by comparing the estimated travel distance and speed with onboard GPS and CAN bus recordings. Figure~\ref{fig:single_tracking} shows the telecom fiber response to the testing vehicle plotted with the y-axis being the fiber distance to virtual sensors (i.e., before geo-localization) or the distance along the road (i.e., after geo-localization). It can be observed that there is an around 100-meter gap between virtual sensors 200 and 300. The gap in the data indicates the lack of fiber-soil coupling due to fiber spooling, which can be corrected after geo-localization, as shown in Figure~\ref{fig:single_tracking} (b). Figure~\ref{fig:tracking_result} shows the ground truth of vehicle location and speed and their estimations before and after geo-localization. Here we can observe the big difference between ground truth and the estimations without geo-localization. To quantify the error, Figure~\ref{fig:tracking_bar} shows the mean absolute errors between ground truth and estimations of vehicle location and speed before and after geo-localization. Our method has a 70$\times$ and an 18$\times$ error rate reduction for the outbound and inbound direction vehicle location estimations, respectively; and a 3$\times$ and a 6$\times$ error rate reduction for the two-direction vehicle speed estimations, respectively.

\begin{figure}[!tb]
    \centering
    \includegraphics[width=0.65\linewidth]{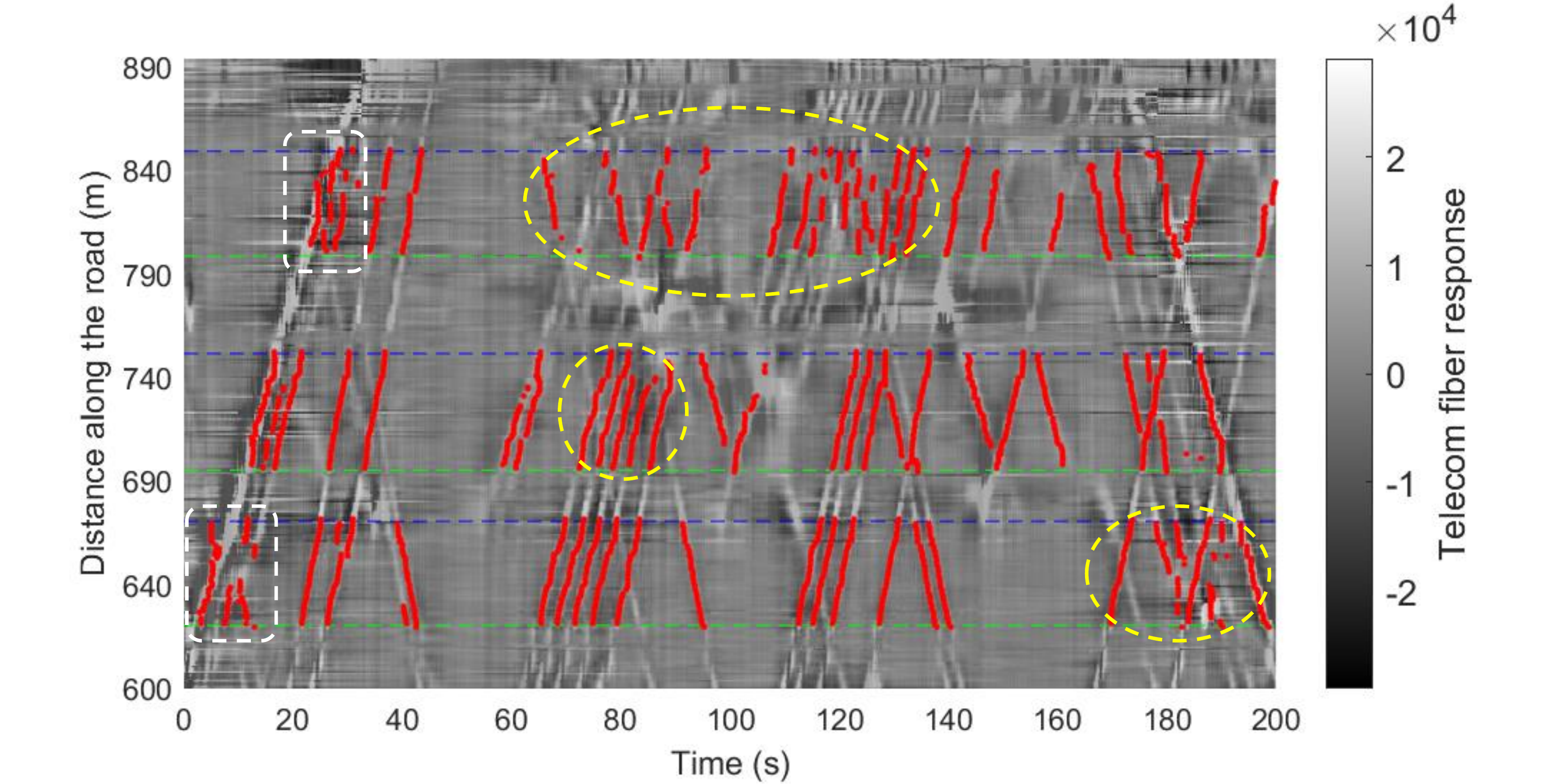}
    \caption{Multiple vehicle tracking. Red curves indicate vehicles tracked by our algorithm. Horizontal dash lines segment road sections by intersections. Dashed line boxes and circles are two types of tracking errors identified and discussed in Section 4.3.}
    \label{fig:multiple_tracking}
\end{figure}

\emph{TelecomTM} performs well for tracking multiple vehicles with daily and busy traffic conditions. Figure~\ref{fig:multiple_tracking} shows an example of our multiple vehicle tracking results. Since vehicle motions in intersections are complex and easy to be overlapping, we segment the road into building blocks divided by intersections along the road, which are indicated by the horizontal dash lines in the figure. We successfully tracked the two-way traffic with various traffic patterns, including multiple vehicles traveling in opposite directions and trains of vehicles passing by in the same direction. Here we also want to discuss two main types of inaccurate tracking results and the reasons causing them. The first type is caused by heavy vehicles marked by white dashed line boxes. When a heavy car is followed by several light vehicles moving across the instrumented road sections, a large quasi-static signal caused by the heavy vehicle masks the quasi-static signals of nearby vehicles, making light vehicles' tracking difficult. We also identified the second type of inaccurate tracking marked by the yellow circles. When multiple vehicles traveling from two directions meet, quasi-static signals created by them have overlapping and more complicated superpositions, causing tracking errors. Especially, there are more missing tracking results in the inbound direction as the traffic is further away from the telecom fiber cable compared to the outbound direction.

\subsection{Vehicle Characterization Results}

\begin{figure*}[!tb]
    \centering
    \begin{subfigure}{0.475\textwidth}
        \centering
        \includegraphics[width=0.6\linewidth]{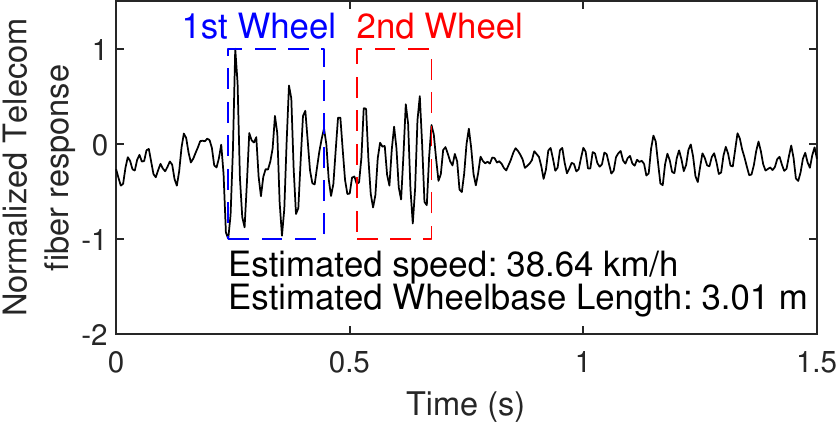}
        \includegraphics[width=0.35\linewidth]{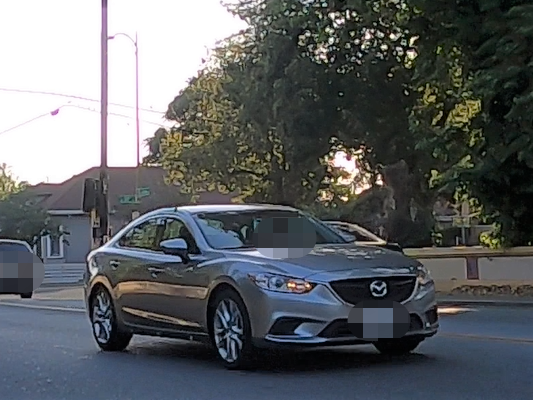}
        \subcaption{}
    \end{subfigure}~
    \begin{subfigure}{0.475\textwidth}
        \centering
        \includegraphics[width=0.6\linewidth]{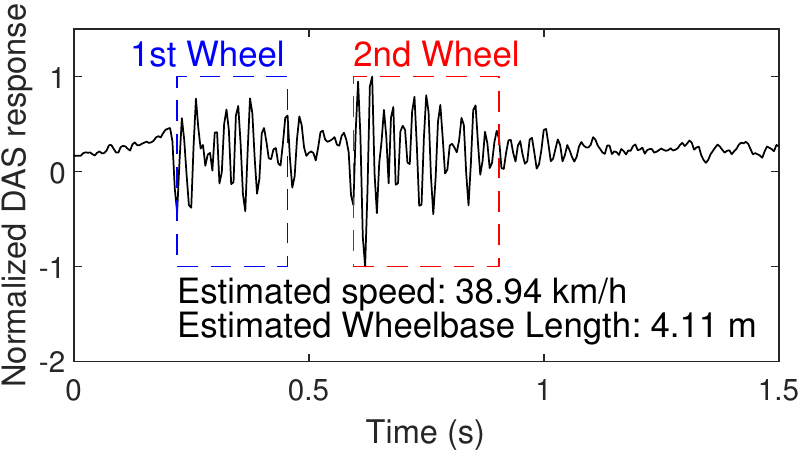}
        \includegraphics[width=0.35\linewidth]{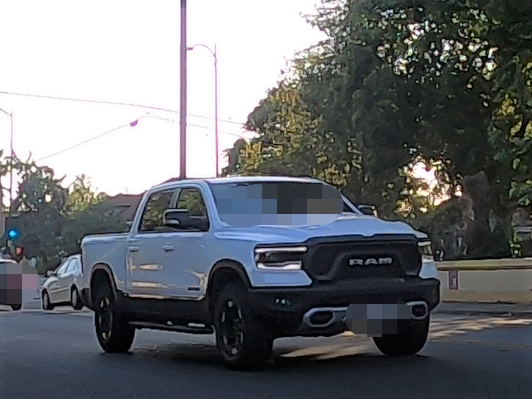}
        \subcaption{}
    \end{subfigure}\\
    \begin{subfigure}{0.475\textwidth}
        \centering
        \includegraphics[width=0.6\linewidth]{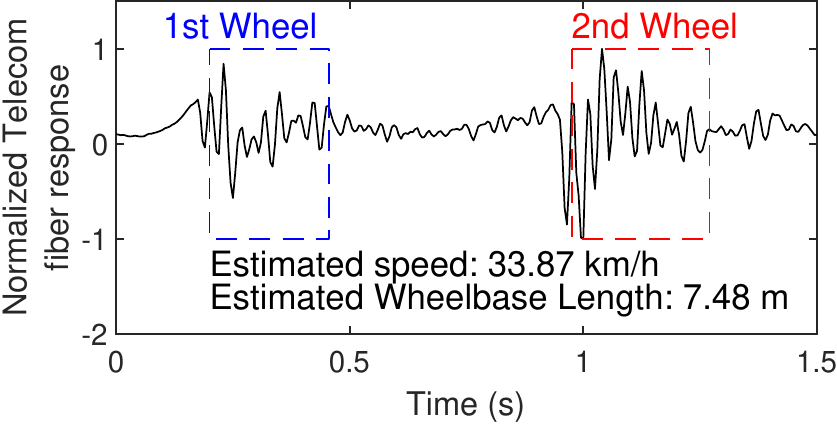}
        \includegraphics[width=0.35\linewidth]{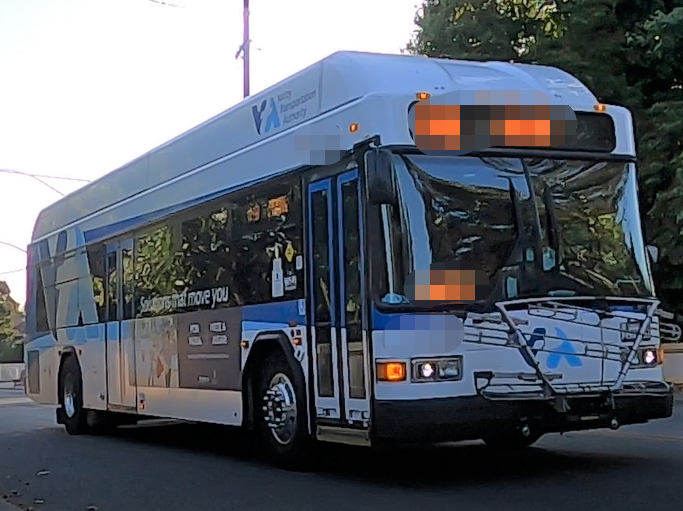}
        \subcaption{}
    \end{subfigure}
    \vspace{-1em}
    \caption{High-frequency responses ($\geq3$ Hz) and the estimated wheelbase lengths of (a) a sedan, (b) a pickup truck, and (c) a bus. Our system estimates wheelbase length by estimating the vehicle speed and time difference between the wheel-induced vibration responses.}
    \label{fig:bus}
\end{figure*}

We evaluated \emph{TelecomTM}'s vehicle characterization performance by comparing our wheelbase and weight estimations with vehicles' specifications. We identified the vehicle models using camera images and searched online for their specifications. Figure~\ref{fig:bus} shows the high-frequency ($> 3 Hz$) responses of three different types of vehicles, a sedan, a pickup truck, and a bus. It can be observed that with a similar moving speed, the larger the vehicle size, the longer the time difference between the wheel-induced vibrations. Overall, our system achieves a $\pm$3.92\% percent error (95\% confidence interval) for wheelbase length estimation, which has a 2$\times$ improvement in error rate reduction (error reduced from $\pm$8\% to $\pm$3.92\%) compared to the commercial piezoelectric pavement sensing system. Furthermore, since the actual weights of vehicles were not allowed to be measured during our experiments, we compared our weight estimation with the vehicles' curb weight which is the weight of the vehicle, including a full tank of fuel and all standard equipment. Although it may not be equal to the actual weight of the moving vehicle, it could reflect our estimation accuracy because there were no heavy trucks during the experiments, and the payloads of the recorded vehicles are considered much smaller than their curb weights. Overall, our system achieves a $\pm$11.98\% percent error (95\% confidence interval) for weight estimation, which has a 3\% improvement in error rate reduction compared to the commercial piezoelectric pavement sensing systems.

\subsection{System Characterization}

In this subsection, we characterize our \emph{TelecomTM} system's performance by examining the effects of sensing distance, vehicle types, crosstalking event, and vehicle moving speed.

\subsubsection{Effect of sensing distance}

Vehicle-induced vibrations attenuate as the distance between the vehicle and fiber cable increases. Traffic in the lane farther from the telecom fiber cable induces smaller responses and has a lower signal-to-noise ratio. To study the effect of sensing distance on vehicle detection and tracking performance, we computed the vehicle detection accuracy, location and speed estimation results for different traffic directions. The outbound traffic travels away from the interrogator, and it is closer to the telecom fiber cable than inbound direction traffic. During our experiments, we recorded 182 inbound vehicles and 206 outbound vehicles. As shown in Figure~\ref{fig:det_result}, detecting vehicles that travel inbound and outbound has similar precision but different recall. The recall for inbound vehicle detection has an up to 45\% accuracy reduction rate compared to that for outbound vehicle detection. Also, Figure~\ref{fig:tracking_bar} shows the MAE of vehicle location and speed estimations for different traffic directions. The location and speed estimations for outbound traffic have a 4.7$\times$ and a 1.5$\times$ improvement in error rate reduction compared to that for inbound traffic. We found that due to the signal-to-noise ratio change, the closer the distance between the traffic and the telecom fiber cable, the more accurate the vehicle detection and tracking performance. {To improve the accuracy for detecting and tracking vehicles in further traffic lanes or roads, we can de-noise the vehicle-induced telecom cable responses to reduce interference among closely traveling vehicles. For instance, works in~\cite{VandenEnde2022a,yuan2022spatial} de-convolve the simulated quasi-static strain response induced by the vehicle from the telecom cable responses to compress the signals into sharp pulses and remove the background noises.}

\subsubsection{Effect of vehicle types}

\begin{figure*}[!tb]
     \centering
    \begin{subfigure}{0.32\textwidth}
        \centering
        \includegraphics[height=3cm]{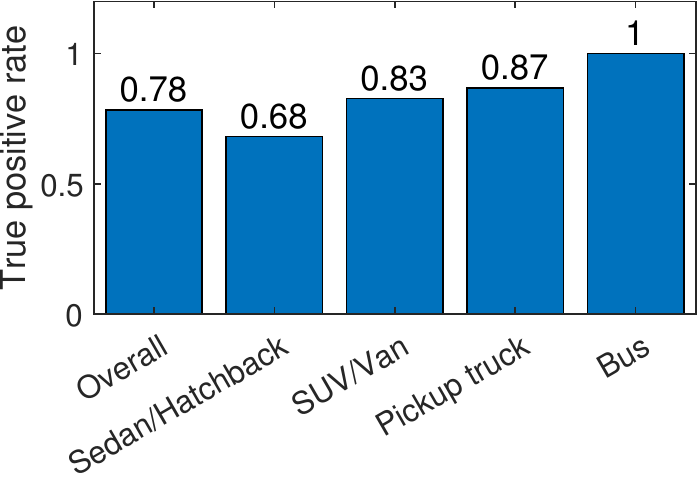}
        \subcaption{}
    \end{subfigure}
    \begin{subfigure}{0.32\textwidth}
        \centering
        \includegraphics[height=3cm]{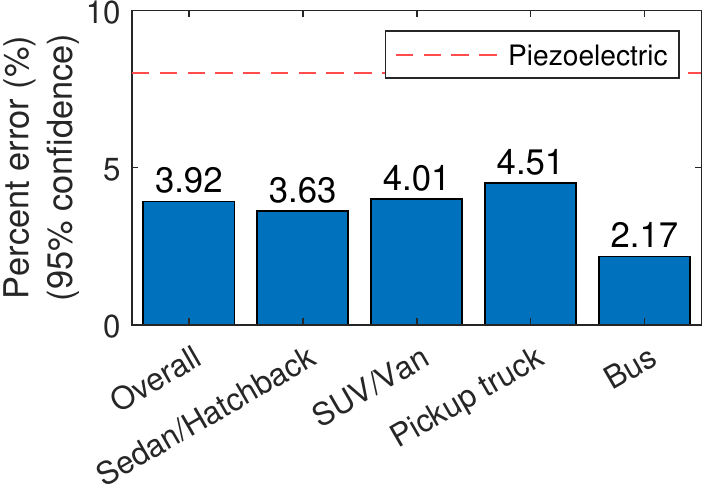}
        \subcaption{}
    \end{subfigure}
    \begin{subfigure}{0.32\textwidth}
        \centering
        \includegraphics[height=3cm]{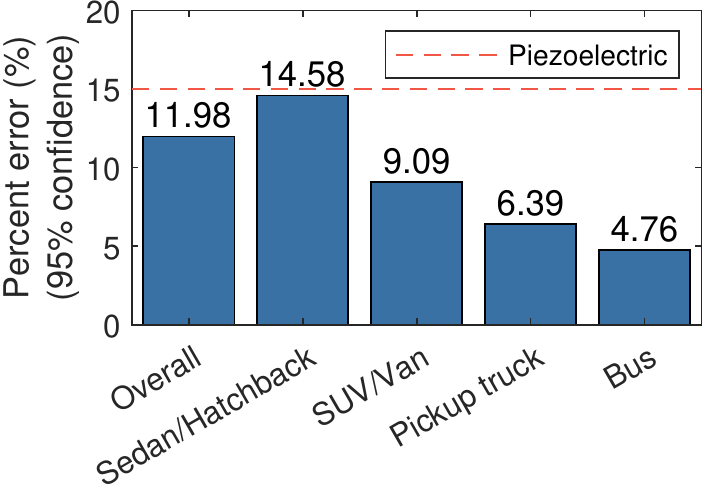}
        \subcaption{}
    \end{subfigure}
    \vspace{-1em}
    \caption{Effect of vehicle types. (a) shows the true positive rate (TPR) of vehicle detection for different vehicle types. The larger the size of a vehicle, the better the vehicle detection result. (b) and (c) show the percent error (95\% confidence interval) for wheelbase and vehicle weight estimations. Red dash lines indicate the percent error of current commercial piezoelectric sensors.}
    \label{fig:char_effect_veh}
\end{figure*}

Vehicles with different sizes, shapes, and weights induce different vibration signals in virtual sensors. We categorize vehicles in our experiments into four types based on their size: 223 sedans/hatchbacks, 114 SUVs/vans, 57 pickup trucks, and 4 buses. Figure~\ref{fig:char_effect_veh} (a) shows the true positive rate (recall) of vehicle detection for different vehicle types moving in the inbound direction. We can observe that the larger the vehicle size, the better the vehicle detection accuracy. It is because larger and heavier vehicles create larger quasi-static responses and have a higher signal-to-noise ratio.

Furthermore, we also studied the effect of vehicle types on vehicle characterization performance. Figure~\ref{fig:char_effect_veh} (b) and (c) show the percent error (95\% confidence interval) of wheelbase length and vehicle weight estimations for different vehicle types. We observe that our system has similar wheelbase estimation errors for different vehicle types. For vehicle weight estimation, {we found that the estimation accuracy increases as the vehicle size and weight increase since heavier vehicles at the same location would induce larger telecom cable response (i.e., higher signal-to-noise ratios)}. 

\subsubsection{Effect of crosstalking events}

\begin{figure}[!tb]
    \centering
    \begin{minipage}{.45\textwidth}
        \centering
        \includegraphics[height=3cm]{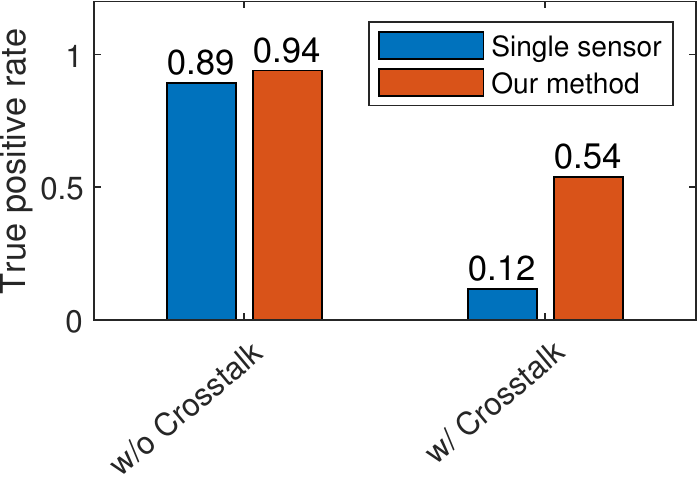}
        \caption{Effect of crosstalking events. True positive rate of vehicle detection with or without crosstalking using a single sensor or our method with distributed sensors.}
        \label{fig:xtalk}
    \end{minipage}\qquad
    \begin{minipage}{0.45\textwidth}
        \centering
        \includegraphics[height=3cm]{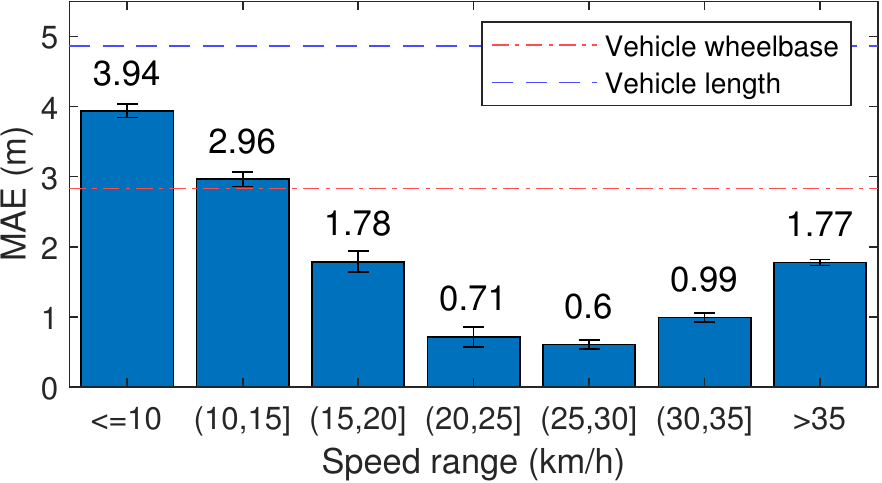}
        \caption{Effect of vehicle moving speed. {Mean absolute value bar chart with a 95\% confidence interval} for vehicle location estimation with different speed ranges. {The number above each bar indicates the mean absolute value.} Red and blue dash lines indicate vehicle wheelbase and length, respectively. Lower and higher speed ranges have larger location estimation errors.}
        \label{fig:effect_speed}
    \end{minipage}
\end{figure}

\emph{TelecomTM} monitors traffic with one-dimensional responses, the axial dynamic strain of the telecom fiber.
Telecom fiber responses induced by multiple vehicles from different lanes passing a virtual sensor at the same time overlap and would be detected as one vehicle event. We define this overlapping as the crosstalking event that mainly affects the inbound (longer sensing distance) vehicle detection performance. During our experiments, 40\% of inbound traffic are crosstalking events. Figure~\ref{fig:xtalk} shows the true positive rate of vehicle detection with or without the crosstalking effect using a single sensor or using our \emph{TelecomTM} method. We can observe that for detection using a single sensor, crosstalking events reduce the true positive rate by 77\%. By leveraging the results from multiple adjacent virtual sensors, we mitigate the issue and improve the TPR from 12\% to 54\% for detecting vehicles having crosstalking.

\subsubsection{Effect of vehicle moving speed}

We studied the effect of vehicle moving speed by investigating the vehicle location estimation errors with respect to different speed ranges, as shown in Figure~\ref{fig:effect_speed}. Here we see that lower and higher speeds have large MAE for vehicle location estimation. Vehicle moving at speed between 25 and 30 km/h has the smallest MAE, 0.6 m. {The larger error for tracking slower vehicles would be because, with the 10-meter gauge length, the vehicle-induced vibration signals are smoothed out, which affects the sensitivity of peak detection-based arrival time estimation for locating slower vehicles. The larger error for tracking faster vehicles would be because these vehicles create larger dynamic responses that reduce the signal-to-noise ratio.} Horizontal dash lines in the figure indicate the wheelbase and total length of the testing sedan. Here we see that the location estimations for all the speed ranges have smaller errors than the vehicle length, and the estimations for the speed $> 15$ km/h have smaller errors than the wheelbase length. Our method has enough location estimation resolution for tracking small vehicles, such as sedans, and for estimating wheelbases with high vehicle speed.

\begin{figure}[!tb]
    \centering
    \includegraphics[width=0.4\linewidth]{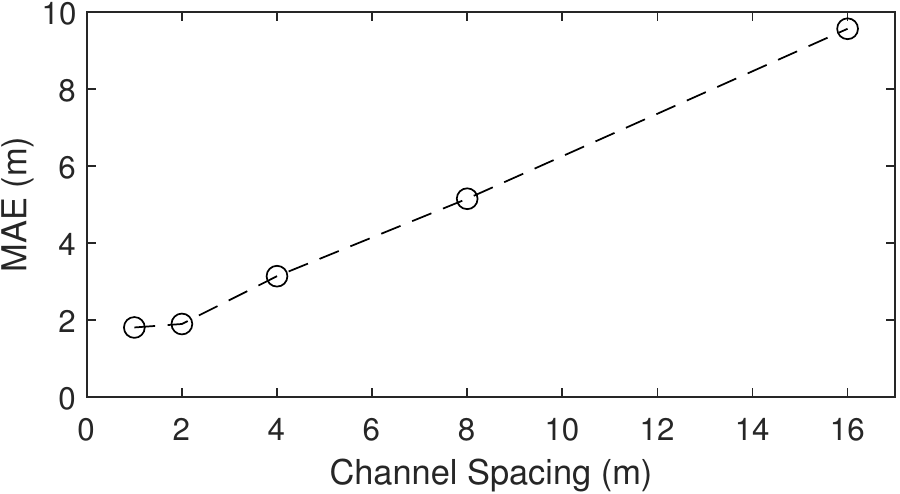}
    \caption{Effect of channel spacing. Multiple vehicle tracking. Mean absolute errors of vehicle location estimation with different channel spacing values.}
    \label{fig:channelspacing}
\end{figure}

{\subsubsection{Effect of channel spacing}

We studied the effect of DAS channel spacing by calculating the mean absolute errors of vehicle location estimation with different channel spacing values, as shown in Figure~\ref{fig:channelspacing}. We can observe that the vehicle tracking error increases as the channel spacing increases. In our work, we first set the channel spacing to be 1 meter, which is the finest spatial interval we can achieve. Then, we chose the gauge length that makes sure vehicle-induced telecom vibrations are clearly visible in the DAS signals.}

\section{Related Work}
{This section provides a review of related works in roadway traffic monitoring, including the state-of-the-art of traffic sensing technologies and existing studies for traffic monitoring using pre-existing telecom fiber cables.

\subsection{The state-of-the-art of traffic sensing technologies}
Much work has been done on traffic monitoring with sensing technologies, including traffic volume estimation, speed estimation, vehicle classification, and weight estimation. The most common technologies are pavement sensing systems, vision-based sensing systems~\cite{5309837,7458203,REINARTZ2006149}, and crowd-sensing systems~\cite{7079458,doi:10.1080/01441640500361108,zhong2021metatp}. Pavement sensing systems, such as inductive loops~\cite{jain2019review,6957957,doi:10.3141/1719-14}, piezoelectric sensors~\cite{zhang2015new,li2006application,jain2019review}, and fiber optic sensors~\cite{tekinay2022applications,yuksel2020implementation}, capture traffic information through measuring the changes on the pavement, such as inductance of the coil, pressure fluctuation, vibration, etc. Although such pavement sensing systems have been used for various traffic monitoring applications, their installation and maintenance require complicated field works that interrupt regular traffic~\cite{bernas2018survey}. On the other hand, installing cameras to monitor road networks is cheaper and less disruptive than pavement sensing systems. Also, many cameras are already installed on roadways for surveillance purposes~\cite{setchell1998applications}. However, such vision-based systems have line-of-sight restrictions, are sensitive to weather conditions, and bring privacy concerns. Crowd-sensing approaches that use cell phone data or Vehicle-to-everything (V2X) data to enable low-cost and efficient traffic monitoring by detecting the drivers'/passengers' phones or the vehicles in motion. However, such crowd-sensing systems cannot provide vehicle characterization information and raise privacy concerns. In summary, the drawbacks of existing traffic monitoring systems hamper their effectiveness in achieving fine-grained and ubiquitous traffic monitoring.
}

\subsection{Traffic monitoring using pre-existing telecom fiber cables}
Existing studies have explored traffic monitoring using telecom fiber cable in three main aspects, including 1) human mobility characterization, 2) vehicle detection and counting, and preliminary work in 3) vehicle speed estimation. 

For {human mobility characterization}, previous studies have demonstrated the strong correlation between the telecom fiber signal variations and the intensity of human activities, including footsteps, traffic, and construction activities~\cite{Lindsey2020,Shen2021}. When there are fewer activities on the sensing site (e.g., during the COVID-19 or spring break of a school), a decrease of amplitude in low-frequency signals is observed. These studies validate the trend by comparing the ground truth mobility data with the seismic noise level (indicated by the signal energy in low-frequency bands) from telecom fiber signals. 

For {vehicle detection and counting}, previous studies have introduced template-matching algorithm and deep deconvolution models to capture the telecom fiber signal variations induced by drive-by vehicles~\citep{Lindsey2020,Ende2021,wang2021ground,VandenEnde2022a,yuksel2020implementation}. These models capture the amplitude changes in the signals to predict the time when the vehicle presents. The template-matching algorithm detects vehicles by comparing the signal with the standard pattern of a signal as the vehicle occurs. The deep deconvolution model deconvolves the vehicle impulse responses from the quasi-static distributed acoustic sensing recordings. Then, a beamforming algorithm is applied to the deconvolved signal rather than the original DAS signal shows improvements in terms of the resolution in vehicle speed estimation, and the detection accuracy.

In addition, preliminary work for {vehicle speed estimation} estimates the speed of a single-vehicle based on the time interval between two adjacent locations of the same detected vehicle~\cite{Yuan2021}. This preliminary work characterizes the traffic-induced surface waves recorded by telecom fiber and compares it with vehicle onboard sensors to validate the vehicle observation in telecom fiber signals. 

While the previous work has validated the feasibility of using pre-existing telecom fiber-optic cables to monitor traffic, these methods are exploratory and mainly focus on coarse-grained traffic flow and speed estimation. In our work, we introduce a systematic approach for traffic monitoring using telecom fiber, which provides a quantitative analysis of traffic data for accurate and fine-grained vehicle detection, tracking, and characterization.

\section{Conclusion and Future works}

In conclusion, we introduced the \emph{TelecomTM} system that uses pre-existing roadside telecom fiber cables to achieve fine-grained and ubiquitous traffic monitoring at low cost with low maintenance. \emph{TelecomTM} uses the distributed acoustic sensing technique to turn telecom fiber cables into virtual strain sensors in a meter-scale spatial resolution. It senses vehicle-induced near-surface vibrations to detect and track vehicles and estimate vehicle positions, speeds, wheelbase lengths, and weights. To overcome the unknown and heterogeneous sensor properties challenge of using this non-dedicated sensing system, we first estimated the geographic location and analyzed the signal pattern of each virtual sensor by matching a testing vehicle's position and quasi-static signals in the telecom fiber induced by its motion. Further, to overcome the challenge of large and complex noise conditions, we developed a spatial-domain Bayesian filtering and smoothing algorithm that estimates the posterior probability of vehicle arrival time at each virtual sensor. Our system can accurately track vehicle motion by converting the estimated arrival time into vehicle positions and speeds. Vehicle wheelbase and weight are estimated by analyzing the time difference between vehicle wheel-induced responses and the quasi-static strain.

We evaluated our system through real-world experiments and extensively characterized its performance with different traffic conditions, vehicle types, sensing distance, and vehicle speeds. Our system detects two-way traffic with 90.18\% accuracy, tracks vehicles with 5.16-meter position estimation error and 3.57 km/h speed estimation error, estimates vehicle wheelbase length with $\pm$3.92\% percent error, and estimates vehicle weight with $\pm$11.98\% percent error.

Our vision of future works {to generalize TelecomTM to multiple roadways with various environmental and operational conditions} is
\begin{itemize}
    \item[1)] {\bf Understanding Environmental Influences:} Environmental influences, including the temperature, humidity, and human activity disturbances on the road surfaces, vary over time and significantly affect the resultant telecom fiber responses. To understand how these environmental factors influence traffic monitoring performance, we plan to collect data under various situations in the long term. Further, we will explore methods to reduce these influences to improve the robustness of our \emph{TelecomTM} system.
    \item[2)] {\bf Characterizing Noise and Uncertainties:} The noise and uncertainties in the telecom fiber responses come from various sources, including different types of conduits, coupling conditions, and soil properties around the cable. To further improve the signal quality and prediction accuracy, we plan to characterize the noise and uncertainties caused by each source by conducting controlled experiments with different source combinations.
    \item[3)] {\bf Exploring Complex Roadways:} In real-life scenarios, the roadways may have more complex configurations than that of our evaluation site. Such configurations include curved and sloped road sections, multiple lanes, etc. For future work, we will explore the capability of our system on these roadways and develop solutions for more complex traffic settings. 
\end{itemize}

\begin{acks}
This research was supported in part by the UPS Foundation Endowment Fund and the Leavell Fellowship on Sustainable Built Environment at Stanford University. The authors also gratefully acknowledge the City of San Jose, California, and OptaSense, a Luna company.
\end{acks}

\bibliographystyle{ACM-Reference-Format}
\bibliography{bib}

\end{document}